\documentclass[final,authoryear,1p,times]{elsarticle}

\usepackage{amssymb}
\usepackage{amssymb}
\usepackage{amsmath,graphicx,amssymb,stfloats,subfigure,multicol,amsthm}
\usepackage{epsfig,epsf,psfrag,amssymb,amsfonts,latexsym,graphicx,mathrsfs,subfigure}
\usepackage[dvips]{color}
\usepackage{textcomp}
\usepackage{nomencl}
\usepackage{algorithmic}
\usepackage{algorithm}
\usepackage{mathrsfs}
\usepackage{hyperref}
\usepackage[hyphenbreaks]{breakurl}
\newtheorem{thm}{Theorem}
\newtheorem{lem}{Lemma}

\def\beq{\begin{equation}}
\def\eeq{\end{equation}}
\def\bea{\begin{eqnarray}}
\def\eea{\end{eqnarray}}
\def\ba{\begin{array}}
\def\ea{\end{array}}
\def\bitem{\begin{itemize}}
\def\eitem{\end{itemize}}
\def\benum{\begin{enumerate}}
\def\eenum{\end{enumerate}}

\def\edoc{\end{document}}

\def\eg{{\it e.g., \/}}
\def\etal{{\it et al. \/}}

\def\ie{{\it i.e.,\ \/}}

\definecolor{bgrd}{rgb}{1,1,1}
\definecolor{grey}{rgb}{0.9,0.9,0.6}
\definecolor{gray}{rgb}{0.5,0.5,0.5}
\definecolor{dkr}{rgb}{0.6,0.2,0.2}
\definecolor{dkg}{rgb}{0,0.5,0}
\definecolor{dkb}{rgb}{0.0,0.1,0.7}
\definecolor{light-gray}{gray}{0.85}

\makeatletter
\newdimen{\captionwidth}
\long\def\@makecaption#1#2{%
\captionwidth .9\hsize
\vskip 10pt%
\setbox\@tempboxa\hbox{#1: #2}%
  \ifdim \wd\@tempboxa >\captionwidth%
    \setbox\@tempboxa\hbox{#1:\hspace*{.5em}}%
    \hfil\parbox{\captionwidth}{\raggedright\hangindent \wd\@tempboxa%
    \hangafter=1\unhbox\@tempboxa#2}\hfill%
  \else\centerline{\box\@tempboxa}%
  \fi
}
\makeatother

\newcommand{\mbbE}{\mathbb{E}}

\newcommand{\Cmsc}{\mathscr{C}}

\def\Cc{{\cal C}}

\def\Rc{{\cal R}}

\def\Uc{{\cal U}}

\journal{Transportation Research: Part D}
\usepackage{lipsum}
\makeatletter
\def\ps@pprintTitle{%
 \let\@oddhead\@empty
 \let\@evenhead\@empty
 \def\@oddfoot{}%
 \let\@evenfoot\@oddfoot}
\makeatother
\begin{document}

\begin{frontmatter}



\title{Market Dynamics and Indirect Network Effects in \\
Electric Vehicle Diffusion\tnoteref{thanks}}


\author[ECE]{Zhe Yu}
 \ead{zy73@cornell.edu}

 \author[Dyson]{Shanjun Li\corref{cor1}}
 \ead{sl2448@cornell.edu}

 \author[ECE]{Lang Tong}
 \ead{lt35@cornell.edu}

 \address[ECE]{School of Electrical and Computer
Engineering, Cornell University, Ithaca, NY 14853, USA.}
 \address[Dyson]{Dyson School of Applied Economics \& Management, Cornell University, Ithaca, NY 14853, USA.}

\cortext[cor1]{Corresponding author}
\begin{abstract}

The diffusion of electric vehicles (EVs) is studied in a two-sided market framework consisting of EVs on the one side and EV charging stations (EVCSs) on the other.  A sequential game is introduced as a model for the interactions between an EVCS investor and EV consumers. A consumer chooses to purchase an EV or a conventional gasoline alternative based on the upfront costs of purchase, the future operating costs, and the availability of charging stations.  The investor, on the other hand, maximizes his profit by deciding whether to build charging facilities at a set of potential EVCS sites or to defer his investments.

The solution of the sequential game characterizes the EV-EVCS market equilibrium.  The market solution is compared with that of a social planner who invests in EVCSs with the goal of maximizing the social welfare.  It is shown that the market solution underinvests EVCSs, leading to slower EV diffusion.  The effects of subsidies for EV purchase and EVCSs are also considered.

\end{abstract}


\end{frontmatter}


\section{Introduction}
The diffusion of electric vehicles in the United States has had mixed results so far. Annual new EV sales increased nearly 7-fold from about $18,000$ in $2011$ to $119,000$ in $2014$. Yet, the market share of EVs was only about $0.73\%$ in the new vehicle market by 2014 ~\citep{Hybridcars:14}.

The reason behind the growth of the EV market, or the lack of it, is multifaceted.  The growth is driven partially by the increasing awareness of the environmental impacts of gasoline vehicles (GVs), superior designs and performance of some EVs, and, by no small measure, subsidies in the form of tax credits provided by the federal and state governments.  However, the EV industry faces stiff skepticism due to the high purchase cost of EVs, the limited driving range, and the lack of adequate public charging facilities.

A similar trend exists in the deployment of public charging services.  The U.S. has built about 9900 charging stations with about 26000 charging outlets~\citep{DoE:15}, due in part to the direct and indirect investments of federal and local governments. For example, the Department of Energy (DoE) provided $230$ million dollars from 2013 to establish $13,000$ charging stations~\citep{DOE:13}. It has been hoped that such investments will stimulate the EV market, driving its market share toward long-term growth and stability.

\begin{figure}[h!]
\subfigure[\small EV stock per million people.]{
  \begin{minipage}{.45\textwidth}
    \includegraphics[width=1\textwidth]{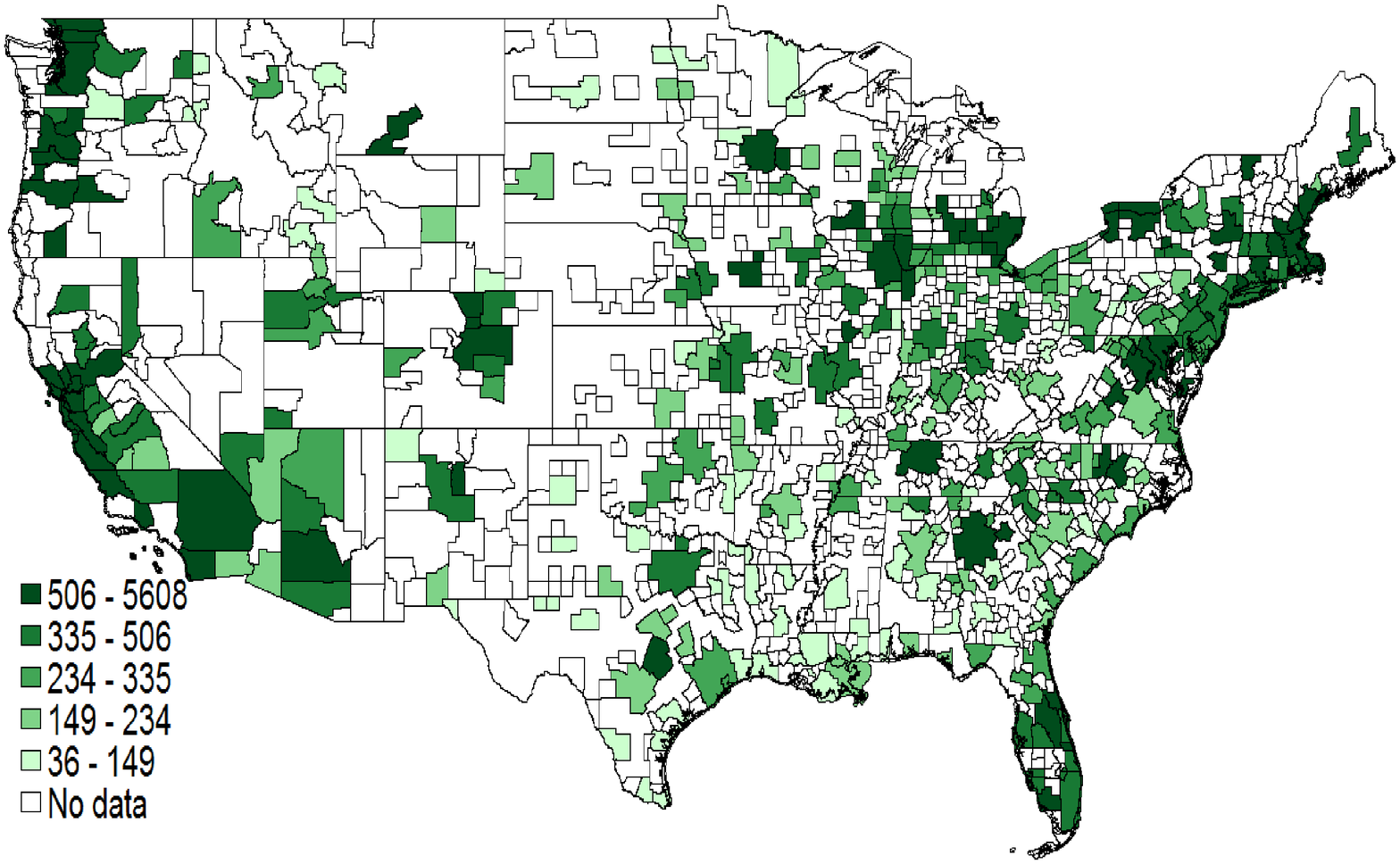}
  \end{minipage}
   }
  \subfigure[\small Public charging stations per million people.]{
  \begin{minipage}{.45\textwidth}
    \includegraphics[width=1\textwidth]{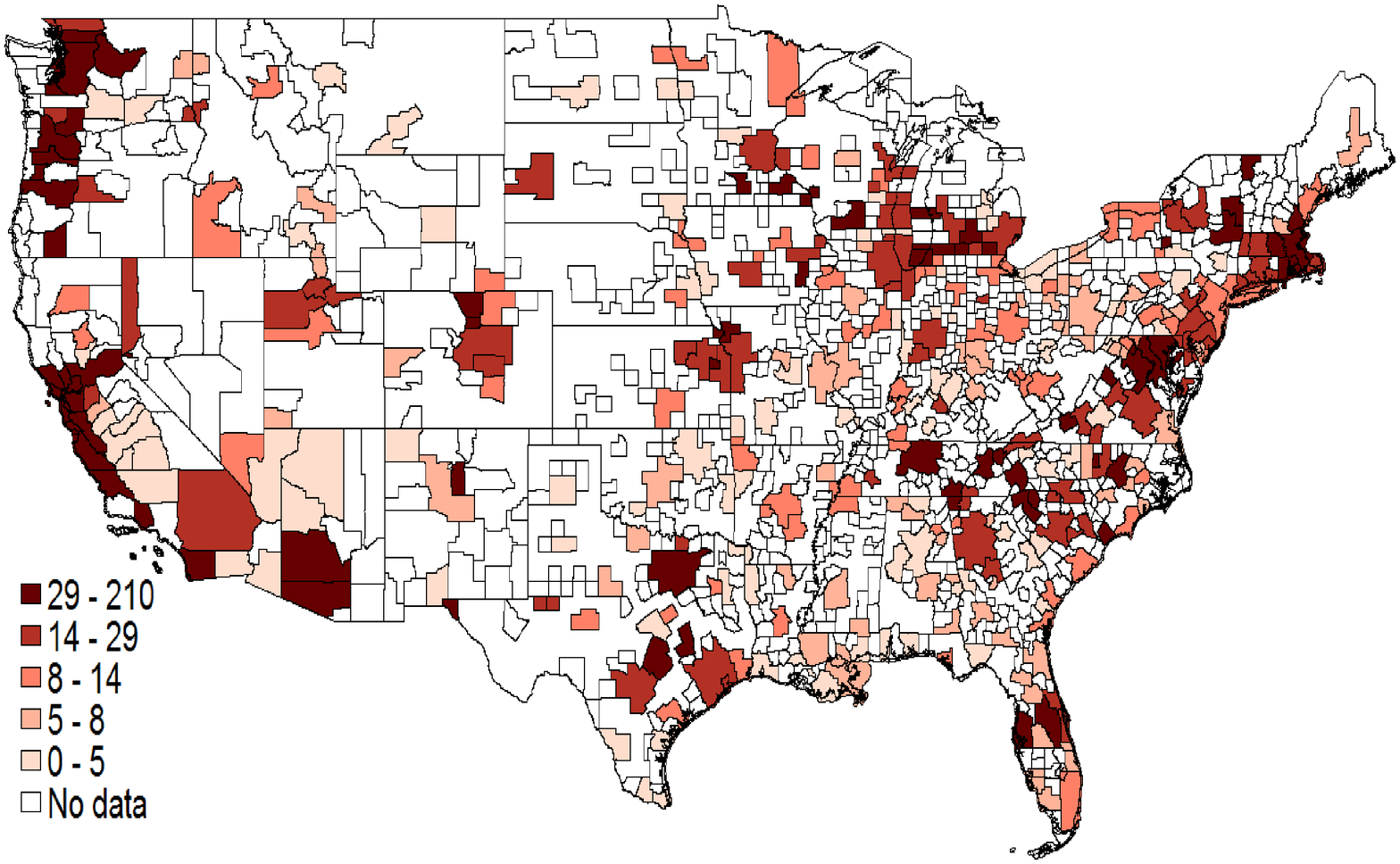}
  \end{minipage}
  }
  \centering
  \caption{EVs and public charging stations in Metropolitan Statistical Areas in 2014~\citep{Hybridcars:14}.}  \label{fig:EV&EVCS}
\end{figure}

The growth trends of EVs and EVCSs have strong temporal and geographic couplings as shown in Fig.~\ref{fig:EV&EVCS}. This could be a manifestation of a cross network effect: consumers' EV adoption in the EV market is affected by the availability of EV charging stations whereas the level of EVCS investment strongly depends on the size of EV stock.

\subsection{Summary of results}
The main contribution of this work is to provide an analytical study on indirect network effects in the EV market. In particular, we formulate a sequential game for the two-sided market and address analytically and numerically the following questions: how does the EV adoption interact with EVCS investment?  How do indirect network effects affect market dynamics? What are the implications of indirect network effects on the public policy?

We introduce a perfect and complete sequential game model for the two-sided EV-EVCS market with an investor as the leader and each consumer a follower.  Through profit maximization, the investor decides whether to build EVCSs at sites chosen (optimally) from a list of candidate sites or to defer his investment with earned interest.  The candidate sites are heterogeneous; each site may have a different favorability rating and different capital costs. The optimal investment decision also includes the optimal pricing of charging services.

Observing investor's decision which defines the locations of EVCSs and the charging prices, a consumer decides whether to purchase an EV or a gasoline alternative. If the choice is an EV, the consumer also decides the preferred charging service.

We provide a solution to the sequential game that includes the optimal decisions for the consumers and the investor. A random utility maximization (RUM) model~\citep{Marschak:60SMMSS} with two different distributions of the consumer vehicle preference is considered.

Under the assumption that the consumer vehicle preference has the type I extreme value distribution, we show that the optimal purchasing decision is a threshold policy on the difference of vehicle preferences of the consumer. A closed-form expression for the EV market share $\eta_e$ is obtained, where $\eta_e$ is a function of the EV purchasing price, the investor's decisions on the number/locations of charging stations, and the charging prices at those locations. The obtained closed-form solution allows us to examine how the investor's decisions and the cost of EVs affect the overall EV market share.

To obtain the optimal investment decision, we first study the optimal operation decision of the investor by fixing the set of EVCS sites. We show that the optimal pricing for EV charging at these sites is such that profits generated from these sites are equal.  We show further that the optimal pricing increases logarithmically with the density of EV charging sites.

The optimal decision on choosing which EVCS sites to build (or deferring investment) is more complicated and is combinatorial in nature.  We provide a greedy heuristic and show that the heuristic is asymptotically optimal as the density of EVCS sites increases.

Under the assumption of the uniform vehicle preference, similar results are also obtained. The optimal purchasing decision of a consumer is again a threshold policy on the vehicle preference.  There is, however, a {\em dead zone effect} of the EVCS density. Specifically, when the density of EVCSs is lower than some threshold, the EV market share is zero.

\subsection{Related works and organization}
There is an extensive literature on two-sided markets and indirect network effects for various products; see \eg the compact disc (CD) player and CD title markets~\citep{Gandal&Kende&Rob:00RJE}, the video game console and video game  markets
\citep{Clements&Ohashi:05RIE, Corts&Lederman:09JIO,Zhou:14}, the  hardware and software markets
\citep{Dube&Hitsh&Chintagunta:10MS}, the credit card market \citep{Armstrong&Wright:04MIMEO}, and  the yellow page and advertisement markets \citep{Rysman:04RES}. Rochet and Tirole \citeyearpar{Rochet&Tirole:04MIMEO} firstly proposed a restrictive definition to distinguish between one-sided and two-sided markets in the context of charge per usage. Caillaud and Jullien \citeyearpar{Caillaud&Jullien:03RJE} pointed out that, one side of the market always waits for the action from the other side. It is thus critical for players to take the right move during the initial stages of the product diffusion.

Exploratory researches on consumer choice on vehicles and plug-in electric vehicle (PEV) market have been conducted. Surveys were carried out to study the consumer attributes on conventional vehicles and PEVs in \citep{Kurani&Etal:96TRD, Potoglou&Etal:07TRD, Lebeau&Etal:12TRD, Delang&Cheng:12TRD}. On the other hand, Bunch \etal \citeyearpar{BunchEtal:93TRA} established and estimated a nested multinomial logit model for the clean-fuel vehicles demand. Similar discrete decision model of vehicle is used in \citep{Ewing&Sarigollu:98TRD, Gao&Kitirattragarn:08TRA, He&Etal:12TRD, Hackbarth&Reinhard:13TRD} to study the preference of hybrid electric vehicle.It is pointed out that charging convenience is a major concern when consumers make the purchase but no analytical result of impact of EV market performance on EVCS investment is presented.

There is a growing literature on the EVCS investment from the operation research and engineering perspectives. For example,  the charging station deployment has been formulated as an optimization problem from the social planner's point of view in  \citep{Ge&Etal:11EMEIT, Frade&Etal:12TRB, He&Etal:13TRM, Chen&Kockelman&Khan:13TRB}.   A location competition problem of charging stations is considered in \citep{Bernardo&Borrell&Perdiguero:13}, where a discrete decision model of charging stations similar to this paper is used. Efficient design of large scale charging is presented in \citep{Chen&Tong:12SmartGrid} and the competition of charging operations is considered in \citep{Chen&Mount&Tong:13HICCS}.

The work of Li \etal \citeyearpar{LiEtal} and the current paper represent the first analysis of the two-sided EV and EVCS  market and the related indirect network effects.  The work in
\citep{LiEtal} focuses on an empirical study of indirect network effects whereas the current paper focuses on the theoretical analysis. This paper builds upon and extends the work of \citep{Yu&Li&Tong:14Allerton}. In particular, this paper extends the model in~\citep{Yu&Li&Tong:14Allerton} by allowing (unobserved) vehicle preferences to have a type I extreme value distribution consistent with the discrete choice model~\citep{McFadden:book}. Also new in this paper are the comparison between the market solution and the decisions of the social planner and the effects of subsidies for EV purchase and EVCS investments.

This paper is organized as follows. The structure of the two-sided market and a sequential game model are described in Sec.~\ref{sec:II}. The solution to the game is obtained by backward induction.  In Sec.~\ref{sec:III}, the consumers' model and the optimal decisions are obtained. The investor's model and optimal strategy are presented in Sec.~\ref{sec:IV}, as well as the social welfare optimization. Discussions about different effects of subsidies and the difference between the private market solution and the socially optimal solution are presented in Sec.~\ref{sec:V}. Sec.~\ref{sec:VI} concludes the paper.

\section{A Sequential Game Model} \label{sec:II}
In this section, we formulate the two-sided market as a two-player sequential game model with perfect and complete information. We introduce the basic structure of the EV-EVCS market, define the players of the game,  and specify the decision process.

\subsection{Two-sided market structure}
\begin{figure}[htb]
\centering
\begin{psfrags}
\psfrag{consumers}{{\small Consumers}}
\psfrag{Gas Cars}{{\small Gas Cars}}
\psfrag{platform}{{\small Platform}}
\psfrag{Cars}{{\small Cars}}
\psfrag{Tesla}{{\small Tesla}}
\psfrag{Leaf or Volt}{{\small EV}}
\psfrag{Charging Service market}{{\small Investor' Choice}}
\psfrag{Universal Charging}{{\small Charging Station}}
\psfrag{Battery Swapping}{{\small Battery Swapping}}
\psfrag{Inverstor}{{\small Investor}}
\psfrag{Gas Station}{{\small Bank}}
\includegraphics[width=0.9\textwidth]{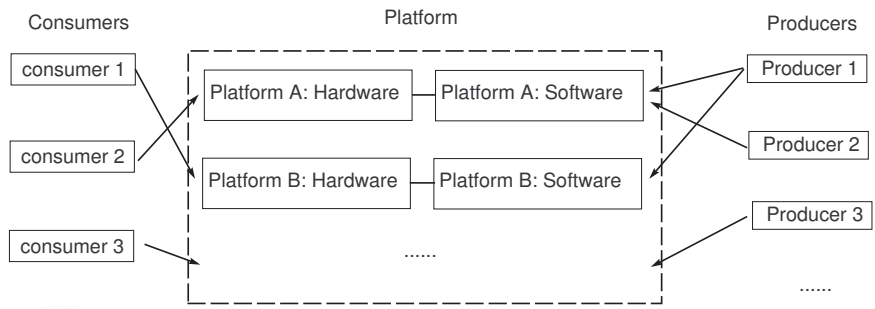}
\end{psfrags}
\caption{\small The structure of a two-sided market.}
\label{fig:twosided}
\end{figure}

A two-sided market typically has a structure as illustrated in Fig.~\ref{fig:twosided}, where we use a generic hardware-software market as an example to describe its basic components. A two-sided market includes a set of platforms, say,  Macbook\texttrademark~and the OS X  operating system as a hardware-software platform by Apple Inc. vs. Dell's Inspiron\texttrademark ~and the Windows 8 as an alternative.

On the one side of the platforms is the consumer who makes her purchase decisions based on her platform preferences, the costs of the platforms, and the available softwares for different platforms.  On the other side of the platforms are the software developers who invest in developing softwares for one particular platform or multiple platforms. The software developer makes his decision based on, among other factors, the costs of developing softwares and the popularity of platforms.

For the two sided EV-EVCS market studied in this paper, we consider two platforms: one is the EV as the ``hardware" and the EVCS as the ``software".  The other is the traditional gasoline vehicle as the hardware and the gas station as the software. On the one side of the platforms are the consumers who decide which type of vehicles to purchase based on the cost of EV, the available charging stations, and the cost of charging. On the other side of the platforms is an investor who decides to build and operate charging stations or to defer his investment and earn interest at a fixed rate\footnote{Because we focus on the early stage of EV diffusion in an environment that gas stations are already well established, the alternative of EVCS investor is not building additional gas stations.}.

\subsection{The investor's decision model}
We assume that the investor is both the builder and the operator of EVCSs. The monopoly assumption is valid at the launch stage of EVCS market since the investment is mainly conducted by the EV manufactories or the government. The investor's decision has two components: the first is an {\em investment decision} on whether to build EVCSs from a list of candidate sites or to defer his investment. The second is an {\em operation decision} on pricing the charging services at those built locations.

Let $\Cmsc=\{s_i=(f_i,c_i), i=1, \cdots, N_L\}$ be the set of candidate sites for charging stations known to the investor.  Site $s_i=(f_i,c_i)$ has two attributes: the favorability rating $f_i$ and the marginal operating cost\footnote{The marginal cost (\$/mile) here is the locational marginal price of wholesale electricity (\$/kWh) normalized by EV efficiency (miles/kWh).} $c_i$. For example, a site at a shopping center may be more attractive than a location that is less frequently visited by consumers.  The capital cost of building and operating at site $s_i$ is denoted by $F(s_i)$.

Given $\Cmsc$ and the utility functions of the consumer, the investor's decision is denoted by $(\Cc, \vec{\rho}) \in 2^\Cmsc \times \Rc^{|\mathcal{C}|}$ where $\Cc \subseteq \Cmsc$ is the set of
  locations selected to build charging stations and $\vec{\rho}=(\rho_1,\cdots,\rho_{|\mathcal{C}|})
  \in\mathcal{R}^{|\mathcal{C}|}$ the vector of  charging prices at the built stations.

Assuming the consumer maximizes her utility, the investor chooses the investment sites and charging prices to maximize the investment profit within his budget $B$. The investment optimization is stated as
\begin{equation}\label{eq:opt_inv}
\begin{array}{ll}
\max_{\mathcal{C},\vec{\rho}}&\Pi(\mathcal{C},\vec{\rho})-\sum_{i=1}^{|\mathcal{C}|}F(s_i)\\
\mbox{subject to}& \sum_{i=1}^{|\mathcal{C}|}F(s_i)\le B
\end{array}
\end{equation}
where $\Pi$ is the operational profit and $F(s_i)$ the building cost of station $i$.

\subsection{The consumer's decision model}\label{sec:IIconsumerModel}
A consumer observes the investor's decision on the location set of charging stations ${\mathcal{C}=\{s_1,\cdots, s_{N_E}\}}$ and the charging price vector $\vec{\rho}=(\rho_1,\cdots, \rho_{N_E})$ where $N_E$ is the number of charging stations. The consumer first chooses the type of  vehicle to purchase. If the choice is an EV,  the consumer also decides on the location of charging.  The action of the consumer is given by $\{\nu,k\}$ where $\nu\in\{E,G\}$ is the vehicle choice (either EV or GV) and  $k\in\{0,1,\cdots,N_E\}$ the preferred charging station. We include $k=0$ for the home charging option. The consumer chooses $\{\nu,k\}$ by maximizing the overall vehicle utility that includes the charging  utility for the EV purchase.

For the vehicle choice, we assume a widely adopted discrete choice model with random utility functions~\citep{Bernardo&Borrell&Perdiguero:13}. The consumer utility model of purchasing a vehicle is assumed as follows.
\begin{equation}
    \begin{array}{c}\label{eq:cutility}
    V_E=\beta_1\mathbb{E}(U_{E})-\beta_2p_E+\Phi+\epsilon_E
   \\
    V_G=\beta_1 \mathbb{E}(U_G)-\beta_2p_G+\Phi+\epsilon_G
      \end{array}
    \end{equation}
    where $U_E$ is the (random) charging utility of consumer's best choice defined in (\ref{eq:csurplus}), $\mathbb{E}(U_{E})$ the expected maximum charging utility, $p_E$ the price of an EV,  $\Phi$ the utility of owning a vehicle, and  $\epsilon_E$ a random vehicle preference of EV.  Variables $\mbbE(U_G)$, $p_G$, and $\epsilon_G$ are similarly defined for the gasoline vehicle.  The consumer's decision is then defined by
     \begin{equation}\label{eq:c_vmax}
    \max \{V_E,V_G\}.
    \end{equation}

The optimization of consumer's vehicle decision also includes optimally choosing charging stations. Specifically, the consumer charging utility at station $i$ is assumed to be random in the following form.
\begin{equation}
U_{i}=\alpha_1 f_i-\alpha_2\rho_i+\epsilon_{i}, i=0,\dots,N_E
\end{equation}
where $f_i$ is the favorability rating,  $\rho_i$ the charging price determined by the investor,  $\epsilon_i$ the random preference of charging station $i$.

Given the realization of the charging preference, ${\vec{\epsilon}=(\epsilon_0,\dots,\epsilon_{N_E})}$, the EV owner chooses charging station ${k\in\{0,1,\dots,N_E\}}$ to maximize her charging utility, \ie
\begin{equation} \label{eq:csurplus}
U_E = \max_{k\in\{0,\dots,N_E\}}U_k.
\end{equation}

For the option of choosing a gasoline vehicle, the number/locations of gas stations will not change with the investor's decisions. Thus the expected maximum fueling utility, $\mbbE(U_G)$, is a constant.

\subsection{The sequential game model}
The sequential game structure of the two-sided EV-EVCS market is summarized as follows.
\bitem
\item The investor's decision is defined by the optimization in (\ref{eq:opt_inv}).  Specifically,
given the set of locations, $\Cmsc$, the investor decides to invest (build and operate) charging stations at a subset ${\mathcal{C} \subseteq \Cmsc}$ and determines the charging price $\vec{\rho}$.  When $\mathcal{C}=\emptyset$, the investor defers his investment and earns interest at a fixed rate.
\item The consumer's decision is defined by (\ref{eq:c_vmax}-\ref{eq:csurplus}).  Specifically,
 having observed the investor's decision,
 $\{\mathcal{C},\vec{\rho}\}$, the consumer chooses $\nu\in\{E,G\}$.  If $\nu=E$, the consumer also chooses charging stations to charge by maximizing her charging utility.
\eitem

The dynamic game is solved by backward induction.  In particular, we first consider the consumer's decision by fixing the investor's choice of charging locations and charging prices.  The optimal consumer's decision is given in Sec.~\ref{sec:III}.    In Sec.~\ref{sec:IV},
the optimal investor's decision is presented.

\nomenclature{$\Pi$}{The operational profit of the investor.}
\nomenclature{$\nu$}{The vehicle choice.}
\nomenclature{$F(s_i)$}{The building cost of station $i$.}
\nomenclature{$B$}{The total budget of the investor.}
\nomenclature{$\Cmsc$}{The set of candidate EVCS sites.}
\nomenclature{$s_i$}{Site $i$.}
\nomenclature{$f_i$}{The favorability rating of site $i$.}
\nomenclature{$c_i$}{The marginal operating cost of site $i$.}
\nomenclature{$N_L$}{The number of candidate sites.}
\nomenclature{$\vec{\rho}$}{The vector of charging price.}
\nomenclature{$\rho_i$}{The charging price at charging station $i$.}
\nomenclature{$\Cc$}{The set of built charging stations.}
\nomenclature{$U_E$}{Charging utility of consumer's best choice.}
\nomenclature{$U_G$}{Refueling utility of consumer's best choice.}
\nomenclature{$U_i$}{Charging utility of charging station $i$.}
\nomenclature{$p_E$}{Price of EV.}
\nomenclature{$p_G$}{Price of GV.}
\nomenclature{$\Phi$}{Utility of owning a vehicle.}
\nomenclature{$\epsilon_E$}{Vehicle preference of EV.}
\nomenclature{$\epsilon_G$}{Vehicle preference of GV.}
\nomenclature{$\epsilon_i$}{Charging station preference.}

\section{Consumer Decisions}\label{sec:III}

\subsection{Consumer Decision Model and Assumptions}

We first summarize the assumptions on the consumer model given in Sec~\ref{sec:IIconsumerModel}.
\bitem
\item[A1.] Consumers are identical and their decisions are statistically independent. Without loss of generality, we focus on the decision of a  single consumer.
\item[A2.] The average charging demand is normalized to $1$.
\item[A3.] The random preference of charging station $i$, $\epsilon_i$, is independent and identically distributed (IID) and follows the type I extreme value  distribution with the probability density function (PDF)
     \[
    f(\epsilon)=e^{-\epsilon}e^{-e^{-\epsilon}}.
    \]
    \item[A4.] The random preference of vehicles $\epsilon_E$ and $\epsilon_G$ are statistically independent.
\eitem

The type I extreme value  distribution is widely used in the discrete choice model. McFadden firstly introduced it in the consumer choice theory and showed it leads to the multinomial logit distribution across choices \citep{McFadden:book}.

\subsection{Consumer Decisions and EV Market Share}

The main result in this section is the structure of the optimal vehicle decision and the characterization of the EV market share as shown in the following theorem.

\begin{thm}[Consumer choice and EV market share]\label{thm:consumerChoice}
\begin{enumerate}
\item If the vehicle preferences $\epsilon_E$ and $\epsilon_G$ follow the type I extreme value distribution, the optimal consumer decision is a threshold policy on the difference of the vehicle preferences $\epsilon_E-\epsilon_G$:
\[
\left\{\begin{array}{ll}
\epsilon_E-\epsilon_G \ge  \tau_e & \mbox{purchase electric vehicles}\\
\epsilon_E-\epsilon_G < \tau_e & \mbox{purchase gasoline vehicles}
\end{array}
\right.
\]
where
\begin{equation}\label{eq:t*}
\begin{array}{rcl}
\tau_e&=&\beta_1\mathbb{E}(U_G)-\beta_2p_G\\
&&-\beta_1\ln(\sum_{i=0}^{N_E}\exp(\alpha_1f_i-\alpha_2\rho_i))+\beta_2p_E.
\end{array}
\end{equation}

The EV market share is given by

\beq\label{eq:eta_Extreme}
\eta_e=\frac{q^{\beta_1}}{q^{\beta_1}+C},
\eeq
where ${C=\exp(\beta_1\mathbb{E}(U_G)-\beta_2p_G+\beta_2p_E)}$ and ${q=\sum_{i=0}^{N_E}\exp(\alpha_1f_i-\alpha_2\rho_i)}$.
\item If the vehicle preference of EV is uniformly distributed with $\epsilon_E\sim \Uc(0,1)$ and $\epsilon_G=1-\epsilon_E$, the optimal consumer decision is a threshold policy on the realization of the consumer preference $\epsilon_E$,
\[
\left\{\begin{array}{ll}
\epsilon_E \ge  \tau_u & \mbox{purchase electric vehicles}\\
\epsilon_E < \tau_u & \mbox{purchase gasoline vehicles}
\end{array}
\right.
\]

where
\small
\begin{equation}\label{eq:t'*}
\begin{array}{rcl}
\tau_u&=&\Big[[\beta_1 \mbbE(U_G)-\beta_2p_G
-\beta_1\ln(\sum_{i=0}^{{N_E}}\exp(\alpha_1 f_i-\alpha_2{\rho_i}))\\
&&+\beta_2p_E+1]/2\Big]_0^1.
\end{array}
\end{equation}
\normalsize
The EV market share is given by
\beq\label{eq:eta_Uniform}
\eta_u=1-\tau_u.
\eeq
\item Under both assumptions, the charging service market share captured by charging station $i$ is given by
\beq \label{eq:Pi_Extreme}
P_{i}=\frac{\exp(\alpha_1 f_i-\alpha_2\rho_i)}{\sum_{k=0}^{N_E}\exp(\alpha_1 f_k-\alpha_2\rho_k)}\triangleq\frac{q_i}{q}.
\eeq
\end{enumerate}

\end{thm}

\begin{proof}
To derive the optimal consumer vehicle decision from (\ref{eq:cutility}-\ref{eq:c_vmax}), we first compute the expected maximum charging utility from (\ref{eq:csurplus}) using the type I extreme value distribution of $\epsilon_i$.  Specifically,
\begin{equation}
\begin{array}{rl}
\mathbb{E}(U_{E})&=\ln(\sum_{k=0}^{N_E}\exp(\alpha_1 f_k-\alpha_2\rho_k))\\
&\triangleq\ln(\sum_{k=0}^{N_E}q_k)=\ln(q),
\end{array}
\end{equation}
where  $q_k=\exp(\alpha_1 f_k-\alpha_2\rho_k)$.

Next, by substituting $\mbbE(U_E)$ into (\ref{eq:cutility}), the consumer's optimal vehicle choice is given by a threshold policy on $\epsilon_E-\epsilon_G$. In particular, the consumer purchases an EV if
\begin{equation}{\label{eq:preferenceDifference}}
\begin{array}{rl}
\epsilon_E-\epsilon_G &\ge\beta_1\mathbb{E}(U_G)-\beta_2p_G\\
                          &~~-\beta_1\ln(\sum_{i=0}^{N_E}\exp(\alpha_1f_i-\alpha_2\rho_i))+\beta_2p_E.
\end{array}
\end{equation}
Under the assumption of uniform distribution, by substituting $\epsilon_G=1-\epsilon_E$ into (\ref{eq:preferenceDifference}), we have (\ref{eq:t'*}).

From  \citep[chap. 4]{McFadden:book}, the EV market share and the EVCS market share are given by  (\ref{eq:eta_Extreme}), (\ref{eq:eta_Uniform}) and (\ref{eq:Pi_Extreme}).
\end{proof}

With Theorem~\ref{thm:consumerChoice}, we examine the trend of EV market share as a function of the charging station density, the charging price, and the price of EV. In particular, under both preference assumptions, the expressions of EV market share is an increasing and concave function of the number of available charging stations.  This means that the marginal return of building additional charging stations reduces with the number of available charging  stations.

\begin{figure}[h!]
\subfigure[\small Uniform  preference distribution.]{
\label{Fig:EVMarketShare_Uniform}
  \begin{minipage}{.45\textwidth}
    \includegraphics[width=1\textwidth]{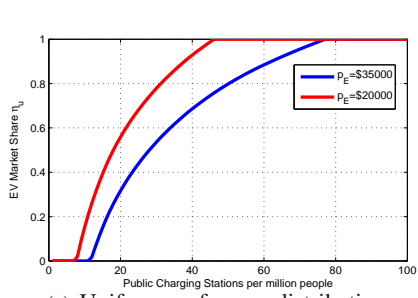}
  \end{minipage}
   }
  \subfigure[\small Type I extreme value preference distribution.]{
  \label{Fig:EVMarketShare_Extreme}
  \begin{minipage}{.45\textwidth}
    \includegraphics[width=1\textwidth]{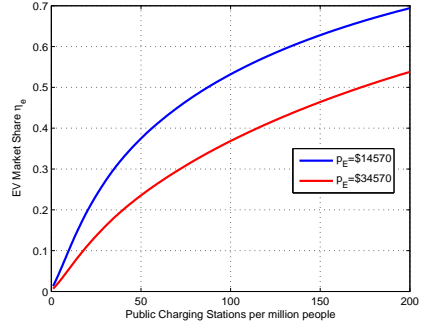}
  \end{minipage}
  }
  \centering
  \caption{ EV market share vs. density of charging stations. $p_G=\$17450$, $\mbbE(U_G)=4.5052$, $\rho_i=0.2\$/\mbox{kWh}$.}
\end{figure}

Fig.~\ref{Fig:EVMarketShare_Uniform} and ~\ref{Fig:EVMarketShare_Extreme} show numerical evaluations of the market share as the density of charging stations.  In addition to the concavity of the market share function, we also observe that the market share accelerates faster with a lower EV purchasing price.

\begin{figure}[h!]
\subfigure[\small Uniform  preference distribution]{
\label{Fig:criticalNum_Uniform}
  \begin{minipage}{.44\textwidth}
    \includegraphics[width=1.15\textwidth]{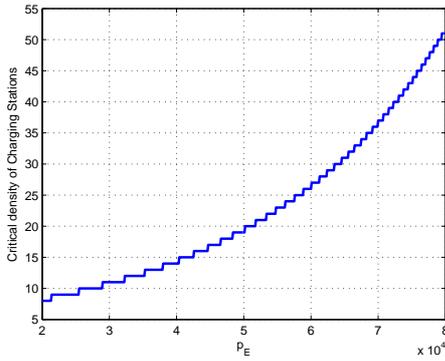}
  \end{minipage}
   }
  \subfigure[\small Type I extreme value preference distribution]{
  \label{Fig:criticalNum_Extreme}
  \begin{minipage}{.44\textwidth}
    \includegraphics[width=1.15\textwidth]{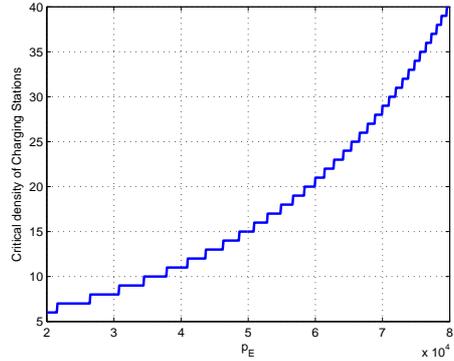}
  \end{minipage}
  }
  \centering
  \caption{Critical density of charging stations vs. EV price $p_E$. $p_G=\$17450$, $\mbbE(U_G)=4.5052$, $\rho_i=0.2\$/\mbox{kWh}$.}
\end{figure}

For the uniform preference model, Fig.~\ref{Fig:EVMarketShare_Uniform} shows a dead zone effect, as a result of the ceiling operation in~(\ref{eq:t'*}).  In particular, the EV market share is zero unless the density of charging stations exceeds a certain level.  In addition, we see that lowering EV purchasing cost helps the market share escaping the dead zone.  Fig.~\ref{Fig:criticalNum_Uniform} shows that the critical density of charging stations at which the market share becomes positive grows as a ``convex'' function of the EV purchasing price. The convexity of this function means that the requirement of initial investment on EVCSs stiffens as the cost of EV purchase increases.

Under the extreme value distribution assumption, there is no dead zone effect. The EV market share $\eta_e$ is always positive. However, if we treat $\eta_e\le5\%$ as a launch failure of EV, there is a critical density below which EV is considered failed. In Fig.~\ref{Fig:criticalNum_Extreme}, the critical density of EVCSs is shown to has a ``convex'' shape in terms of EV prices.

%

\nomenclature{$\tau_e$}{Threshold of vehicle preference under type I extreme value distribution assumption.}
\nomenclature{$\tau_u$}{Threshold of vehicle preference under uniform distribution assumption.}
\nomenclature{$\eta_e$}{EV market share under type I extreme value distribution assumption.}
\nomenclature{$\eta_u$}{EV market share under uniform distribution assumption.} 

\section{Investor Decisions}\label{sec:IV}
After the discussion about the consumer model and her decision, we now focus on the investor decision model that includes the selection of the charging stations locations and the optimal pricing of charging.

\subsection{Investor Decision Model and Assumptions}

We make the following assumptions about the investor model:
\bitem
\item[B1.] We consider a single investor who also operates all charging stations. This implies the monopolistic competition in the charging service market.
\item[B2.] We assume that the deferred investment earns interest at a rate of $\gamma$.
\item[B3.] The investor knows the utility functions of the consumers.
\eitem

To solve the optimization in (\ref{eq:opt_inv}), we proceed with backward induction. In Sec~\ref{sec:optimalPrice}, we  find the optimal pricing with fixed EVCS locations. In Sec~\ref{sec:optimalLocation} we optimally choose the charging station locations.

\subsection{Optimal Charging Price}\label{sec:optimalPrice}

Given the set of charging station locations $\mathcal{C}$, the investor determines the optimal charging price $\vec{\rho}$ to maximize the total operation profit. Specifically, the investor has the following optimization.
\beq
\max_{\vec{\rho}}\Pi=\max_{\vec{\rho}}\eta(\vec{\rho})\sum_{i=1}^{N_E}P_i(\vec{\rho})(\rho_i-c_i),
\eeq
where  $\eta(\vec{\rho})$ is the expected EV market share given in Theorem~\ref{thm:consumerChoice} (here we make the dependency on charging price explicit),  $P_i(\vec{\rho})$ the market share of station $i$, \ie the fraction of EV owners who charge at station $i$, and $c_i$ the marginal operation cost of station $i$. The optimal charging price $\rho^*_i$  is given by the following theorem.

\begin{thm}[Charging price]\label{thm:optimalPrice}

For fixed set of  charging stations $\mathcal{C}=\{(f_i, c_i), i=1,\cdots, N_E\}$,  the optimal charging price $\rho_i^*, i=1,\cdots, N_E$ generates uniform profits across charging stations.  In particular,\\
\begin{enumerate}
\item Under the type I extreme value vehicle preference distribution assumption,
\begin{equation}\label{eqn:optimalPrice_Extreme}
\rho^*_{i,e}-c_i=\frac{1}{\alpha_2\beta_1(1-\eta_e(\vec{\rho}_e^*))(1-P_0(\vec{\rho}_e^*))+\alpha_2P_0(\vec{\rho}_e^*)},
\end{equation}
where $\eta_e$ is the market share of EV, $\vec{\rho}_e^*=(\rho^*_{1,e},\dots,\rho^*_{N_E,e})$ the vector of optimal charging price, $\rho_{0,e}$ the cost of charging at home, and ${P_0(\vec{\rho}_e^*)=\frac{\exp(\alpha_1 f_0-\alpha_2\rho_{0,e})}{\exp(\alpha_1 f_0-\alpha_2\rho_{0,e})+\sum_{k=1}^{N_E}\exp(\alpha_1 f_k-\alpha_2\rho^*_{k,e})}}$  the probability that the consumer charges at home.
\item Under the uniform vehicle preference distribution assumption,
\begin{equation}\label{eqn:optimalPrice_Uniform}
\rho^*_{i,u}-c_i=\frac{1}{\frac{\alpha_2\beta_1(1-P_0(\vec{\rho}_u^*))}{2 \eta_u(\vec{\rho}_u^*)}+\alpha_2P_0(\vec{\rho}_u^*)},
\end{equation}
\end{enumerate}
where $\eta_u$, $\vec{\rho}_u^*$, $\rho_{0,u}$, and $P_0(\vec{\rho}_u^*)$ are similarly defined.
\end{thm}

\begin{proof}
This is a direct consequence of the first order optimality condition.
\end{proof}

Note that the right hand sides of equations (\ref{eqn:optimalPrice_Extreme}, \ref{eqn:optimalPrice_Uniform}) are the same for any charging station $i\in\{1,\dots,N_E\}$. This means that the profits generated from different charging stations are the same. Equations (\ref{eqn:optimalPrice_Extreme}, \ref{eqn:optimalPrice_Uniform}) do not have closed-form solutions but the optimal prices can be solved numerically. Since ${0< P_0(\vec{\rho}_j^*)<1}$ and ${0<\eta_j\le1}$ for either ${j\in\{e,u\}}$, the revenue is strictly positive.

As $N_E\rightarrow\infty$, the public charging of EV becomes more and more convenient, which not only motivates consumers to purchase EVs but also encourages them to charge outside home. As a result, the EV market share $\eta_j\rightarrow1$ and the fraction of charging at home $P_0(\vec{\rho}_j^*)\rightarrow0$. Based on this trend, we have the convergence of the marginal charging profit shown in the following theorem.

\begin{thm}[Charging price convergence]\label{thm:priceConvergence}
Consider a fixed set of charging stations ${\mathcal{C}=\{(f_i, c_i), i=1,\cdots, N_E\}}$.  Let ${v=\sum_{i=1}^{N_E}\exp(\alpha_1f_i-\alpha_2c_i)}$ be the sum of exponentials of systematic charging utilities.
\begin{enumerate}
\item For the extreme distributed vehicle preference, the per-charging station profit ${r_e=\rho^*_{i,e}-c_i}$ grows logarithmically with the sum utilities, \ie
\begin{equation}\label{eqn:priceConvergence_Extreme}
r_e=\ln v/\alpha_2  + o(\ln v).
\end{equation}
\item For uniformly distributed vehicle preference, the charging profit ${r_u=\rho^*_{i,u}-c_i}$ is strictly increasing with the number of charging facilities and converges to a constant. Specifically,
\beq\label{eqn:priceConvergence_Uniform}
r_u=2/\alpha_2\beta_1+o(1)~~(v\rightarrow+\infty).
\eeq
\end{enumerate}
\end{thm}
\begin{proof}
See ~\ref{sec:priceConvergence}.
\end{proof}

In Theorem~\ref{thm:priceConvergence}, the increasing per-charging station profit is because of the assumption of monopolistic investor. More charging stations motivate more consumers to purchase EVs and bring larger charging demand. The investor will take the advantage and set a higher markup.

The profit is also affected by the consumer sensitivity to the charging price. When consumers are more sensitive to the price ($\alpha_2$ is large), the optimal charging price is close to the marginal cost across charging stations.

The different convergence comes from different preference assumptions. Under the extremely distributed preference assumption, the EV market share strictly increases as the density of EVCSs increases. Thus the profit grows logarithmically because of the expanding charging demand. While under the uniform distributed preference assumption, the EV market share reaches the upper limit when there are enough EVCSs. So the profit converges to a constant.
%
%

\nomenclature{$r_j$}{The uniform charging profit under different assumptions.}
\nomenclature{$P_i$}{The market share of charging station $i$.}
\nomenclature{$v_i$}{The exponential systematic utility.}
\nomenclature{$v$}{Sum of the exponential systematic utility.} 

\subsection{Optimal Charging Station Locations}\label{sec:optimalLocation}
After the discussion about the optimal charging price, we consider the choice of charging station locations.
Given the set of location candidates $\Cmsc=\{s_i=(f_i, c_i),~i=1,\cdots, N_L\}$, the investor has the following optimization.

\begin{equation}\label{eqn:optimalInvestment}
\begin{array}{ll}
\max_{{\mathcal C} \subseteq \Cmsc}&  \Pi(\mathcal{C},\vec{\rho}_j^*(\Cc))-\sum_{i=1}^{|\mathcal{C}|}F(s_i)\\
\mbox{subject to}& \sum_{i=1}^{|\mathcal{C}|}F(s_i)\le B
\end{array}
\end{equation}
where $F(s_i)$ is the building cost of charging station $s_i$ and ${\Pi(\mathcal{C},\vec{\rho}_j^*(\Cc))}$ the operational profit.

In general, the optimal investment decision from (\ref{eqn:optimalInvestment}) requires combinatorial search for $\mathcal{C}$, which is computationally inefficient and sometimes not tractable. However, the convergence of the optimal charging prices across charging stations in Theorem~\ref{thm:priceConvergence} makes it possible to separate the price decision and the location choice, which leads to a linear complexity heuristic algorithm.

\begin{algorithm}
\caption{Greedy Investment Algorithm}
\label{alg:investmentDecision}
\begin{algorithmic}
\STATE $1$. Compute the exponential systematic utility ${v_i=\exp (\alpha_1 f_i-\alpha_2c_i)}$  and sorted list $\{v_{(i)}\}$.
\STATE $2$. Set $N=1$.
\WHILE{$N\le N_L$ and $\sum_{i=1}^{N} F(s_i)\le B$}
\item Compute $\tilde{P}_{N}\triangleq \Pi(s_1,\cdots,s_{N})-\sum_{i=1}^{N} F(s_i)$.
\item \IF {$\tilde{P}_{N} < \tilde{P}_{N-1}$ or $\sum_{i=1}^{N} F(s_i)>B$}
\STATE STOP;
\ELSE
\STATE $N\leftarrow(N+1)$.
\ENDIF
\ENDWHILE
\end{algorithmic}
\end{algorithm}

The Greedy Investment Algorithm (GIA) given in Algorithm~\ref{alg:investmentDecision} first ranks the charging stations by the exponential systematic part of the charging utility, ${v_i=\exp(\alpha_1 f_i - \alpha_2c_i)}$.  It then adds charging stations to the investment list one at a time in the decreasing order of exponential systematic utility $v_i$ until either the budget is exhausted or the cumulated profit starts to decrease. 

By ignoring the dependency of charging locations in the marginal charging profit $(\rho^*_{i,j}-c_i)$ in (\ref{eqn:optimalInvestment}), the GIA is not optimal in general.  As $N_E$ increases, however, the marginal charging profit increases and converges, which makes the algorithm asymptotically optimal.

\begin{thm}[Asymptotic optimality]\label{thm:asymptoticOptimal}
Assume the building costs of charging stations are constant, \ie $F(s_i)=(1+\gamma)F_0$, where $\gamma$ is the interest rate. There exists an $M>0$ such that when $N>M$, the greedy algorithm is optimal under both the type I extreme value distribution and the uniform distribution assumption.
\end{thm}

\begin{proof}
See ~\ref{sec:ProofAsy}.
\end{proof}

After obtaining the optimal set of charging stations $\mathcal{C}^*$ and the optimal charging price vector $\vec{\rho}_j^*$, the investor makes the investment if the investment profit $(\Pi(\mathcal{C}^*,\vec{\rho}_j^*)-\sum_{i=1}^{|\mathcal{C}^*|}F(s_i))$ is positive. Otherwise, the investor will defer his investment and earn interest at rate $\gamma$.

To make $(\Pi(\mathcal{C}^*,\vec{\rho}_j^*)-\sum_{i=1}^{|\mathcal{C}^*|}F(s_i))$ positive, the EV price and the building costs of charging stations need to be low enough, which implies that the subsidies for EV purchase and charging stations are necessary to the successful launch of EV.

\subsection{Social welfare optimization}\label{sec:socialWelfare}
We now consider the difference between the solution of the private market defined in Sec.~\ref{sec:optimalLocation} and that of a social planner who makes investment decisions based on social welfare maximization.

Recall the investor utility $S_I(\mathcal{C},\vec{\rho}_j)$ and the consumer utility $S_C(\mathcal{C},\vec{\rho}_j)$ given in (\ref{eq:opt_inv}) and (\ref{eq:c_vmax}):
   \[
    \begin{array}{ll}
    S_I(\mathcal{C},\vec{\rho}_j)&=\Pi(\mathcal{C},\vec{\rho}_j)-\sum_{i=1}^{|\mathcal{C}|}F(s_i),\\
    S_C(\mathcal{C},\vec{\rho}_j)&=\mbbE(\max \{ V_E(\mathcal{C},\vec{\rho}_j,\epsilon_E), V_G(\epsilon_G)\}).
    \end{array}
    \]
    Under the type I extreme value vehicle preference distribution assumption, the consumer utility is stated as
    \[
    \begin{array}{l}
    S_C(\mathcal{C},\vec{\rho}_e)=\ln\Bigg[\Bigg(\sum_{i=0}^{|\mathcal{C}|}\exp(\alpha_1f_i-\alpha_2\rho_{i,e})\Bigg)^{\beta_1}C_1+C_2\Bigg],\\
    C_1=\exp(-\beta_2p_E+\Phi),~C_2=\exp(\beta_1 \mbbE(U_G)-\beta_2p_G+\Phi).
    \end{array}
    \]

Under the uniform vehicle preference distribution assumption, the consumer utility is stated as:
    \[
    S_C(\mathcal{C},\vec{\rho}_u)=\Bigg[\Bigg(\eta_u(\mathcal{C},\vec{\rho}_u)\Bigg)^2+\beta_1 \mbbE(U_G)-\beta_2p_G+\Phi-\frac{1}{2}\Bigg].
    \]

    Assume that the social planner does not determine the charging price or the vehicle price, he only determines the set of charging stations to build. The social planner's decision is stated as:

    \[
    \begin{array}{ll}
    \max_{\mathcal{C}\subseteq\Cmsc}& S_C(\mathcal{C},\vec{\rho}_j^*(\mathcal{C}))+S_I(\mathcal{C},\vec{\rho}_j^*(\mathcal{C}))\\
    \mbox{subject to}&\sum_{i=1}^{|\mathcal{C}|}F(s_i)\le B
     \end{array},
    \]
    where $\vec{\rho}_j^*(\mathcal{C})$ is the vector of the optimal charging prices determined by the charging station operators given the charging station locations.

     The greedy investment algorithm in Table \ref{alg:investmentDecision} can also be applied to solve for the social welfare optimized investment in charging stations.  The following theorem characterizes the difference between the social welfare optimal solution and the market solution.
\begin{thm} [Social welfare]\label{thm:socialWelfare}
Let $\mathcal{C}^*$ be the optimal set of charging stations determined by the investor, and assume $|\mathcal{C}^*|\gg1$. Let $\mathcal{C}^{**}$ be the optimal charging locations determined by the social planner. Under both the type I extreme value distribution and the uniform distribution assumptions, $|\mathcal{C}^{**}|>|\mathcal{C}^*|$.
\end{thm}
\begin{proof}
See ~\ref{proof:socialWelfare}.
\end{proof}

Theorem \ref{thm:socialWelfare} implies that the monopolistic market solution tends to under-build charging stations. The under-provision of EVCSs and lower adoption of EVs relative to the socially optimal outcomes are due to two types of market failures: market power and indirect network effects (or externalities). The assumption of monopolistic investor leads to under-provision of EVCSs and a higher charging price than a competitive solution. This will in turn lead to a lower EV adoption. Therefore, introducing competition in EVCS provision will help EV diffusion. While this form of market failure is a result of our model setup and can be relaxed, the second form of market failure is inherent in the EV market as empirically confirmed in [20]. Indirect network effects are externalities which are not accounted for in individual investment and purchase decisions. They will lead to a wedge in socially optimal outcomes and market outcomes, which justifies government interventions. For example, government can provide subsidies to charging station investors as the U.S. DOE does through various funding programs or mandate the provision of charging stations in real estate development as  recently implemented in China~\citep{Loveday:14}. 
\section{Discussions}\label{sec:V}%
\subsection{Effects of subsidy}
We consider here the effects of subsidy, either to EV consumers or to the investor of EV charging stations. The results obtained in Sec.~\ref{sec:III} and Sec.~\ref{sec:IV} provide the basis for numerical results presented here. In numerical simulations, coefficients from \citep{LiEtal} are used and the type I extreme value vehicle preference distribution case is considered.

Fixing the total policy budget as 230 million dollars, we vary the weight of subsidy for EV purchase among the total policy budget. A bisection algorithm is applied to search for the subsidy amounts for each EV and EVCS so that the constraints of total budget and budget weight are satisfied.

\begin{figure}[h!]
\subfigure[\small EV market share vs. subsidy weight.]{
\label{Fig:msVSRatio}
  \begin{minipage}{.47\textwidth}
    \includegraphics[width=1.17\textwidth]{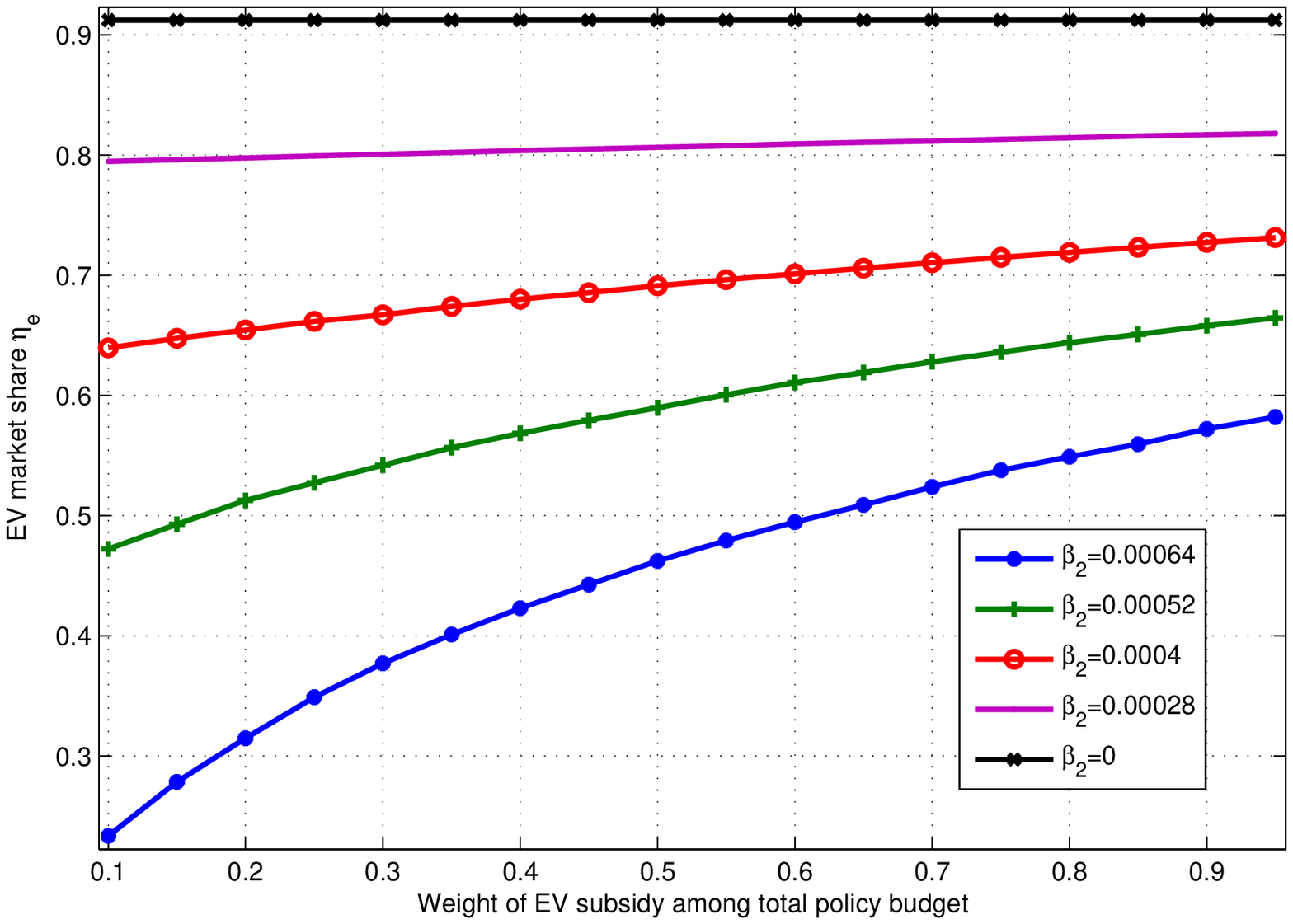}
  \end{minipage}
   }
  \subfigure[\small Density of EVCSs vs. subsidy weight.]{
  \label{Fig:numCSVSRatio}
  \begin{minipage}{.47\textwidth}
    \includegraphics[width=1.15\textwidth]{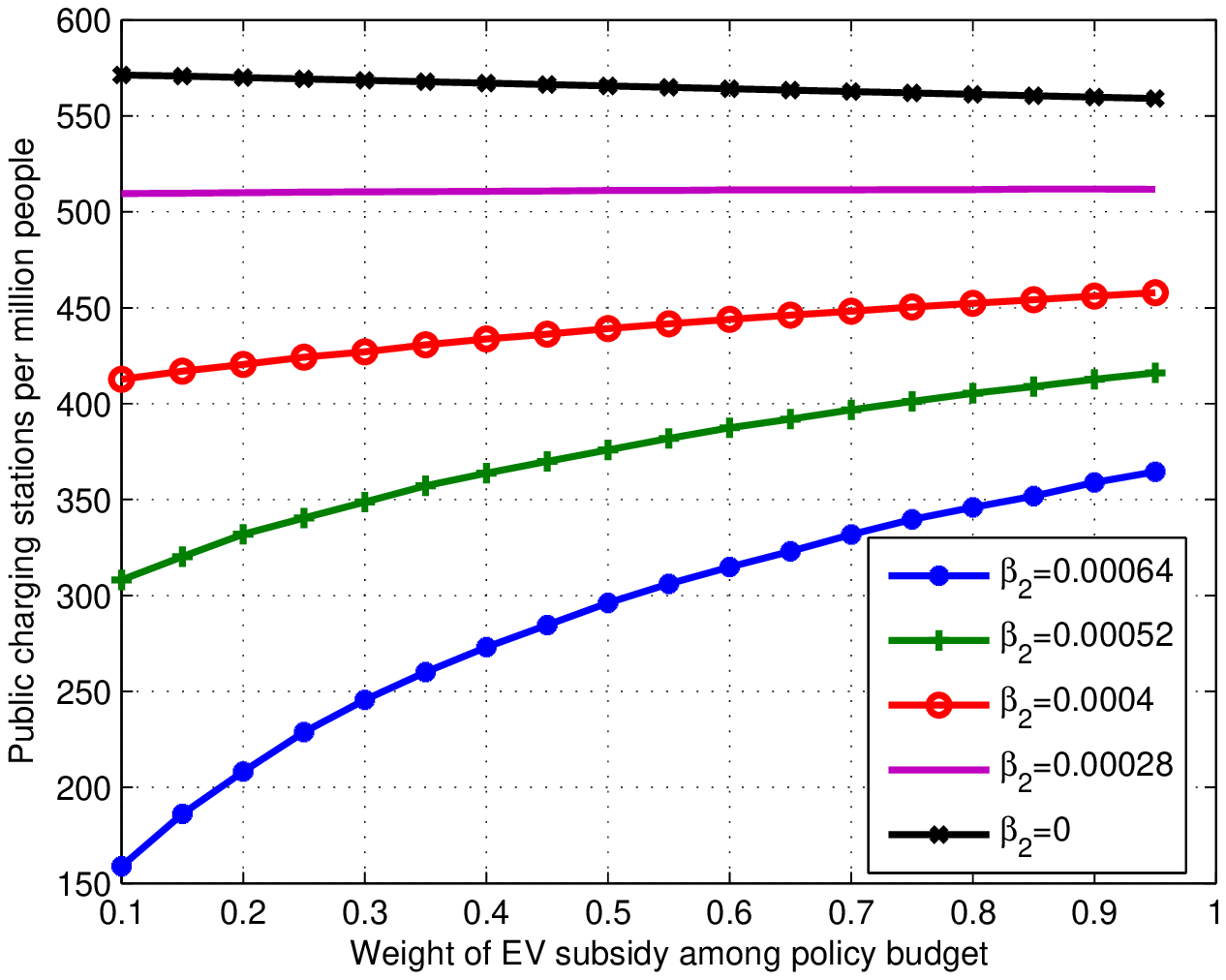}
  \end{minipage}
  }
  \centering
  \caption{Subsidy effect with different coefficients}
\end{figure}

Fig.~\ref{Fig:msVSRatio} shows the EV market share against subsidy weight with different values of $\beta_2$. In the utility model (\ref{eq:cutility}), $\beta_2$ represents the consumer sensitivity to the EV price. When $\beta_2$ is large, consumers care more about the EV price than the characteristics of charging facilities. Increasing EV subsidy dramatically boosts up the EV market share. On the other hand, when $\beta_2$ is close to $0$, consumers mainly concern about the charing services and the subsidy for EV purchase plays a tiny role in the EV market share evolution.

Similar impact exists in the EVCS market. As shown in Fig.~\ref{Fig:numCSVSRatio}, when $\beta_2$ is large, the consumer is more sensitive to EV price. In this case, EV subsidy not only stimulates EV purchasing but also drives the investor to deploy more charging facilities because of the EV popularity. When $\beta_2$ is close to $0$, putting more weight to EV subsidy discourages investment in EVCSs. Fewer charging facilities are invested thus more subsidies for EV purchase, on the contrary, draws down the EV market share.


\subsection{Socially optimal solution vs. private market solution}
In Sec.~\ref{sec:socialWelfare}, the analytical result shows the socially optimal solution requires to invest in more charging facilities than that from the private market solution. In this section, a numerical result is presented to illustrate this difference in the EVCS market.

\begin{figure}[h!]
\subfigure[\small EV market share vs. EV price.]{
\label{Fig:marketShareSW}
  \begin{minipage}{.47\textwidth}
    \includegraphics[width=1.17\textwidth]{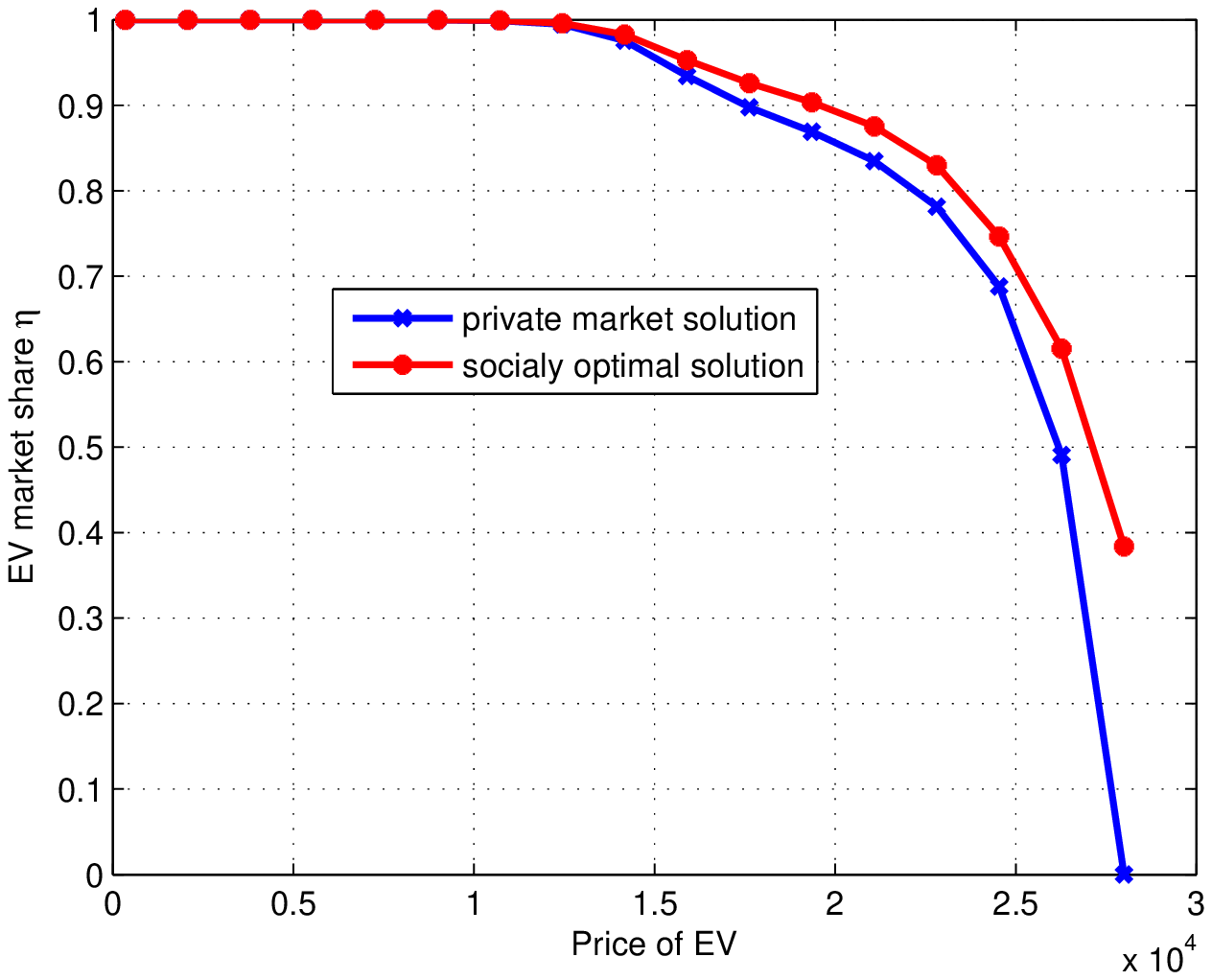}
  \end{minipage}
   }
  \subfigure[\small Density of EVCS vs. EV price.]{
  \label{Fig:densitySW}
  \begin{minipage}{.47\textwidth}
    \includegraphics[width=1.15\textwidth]{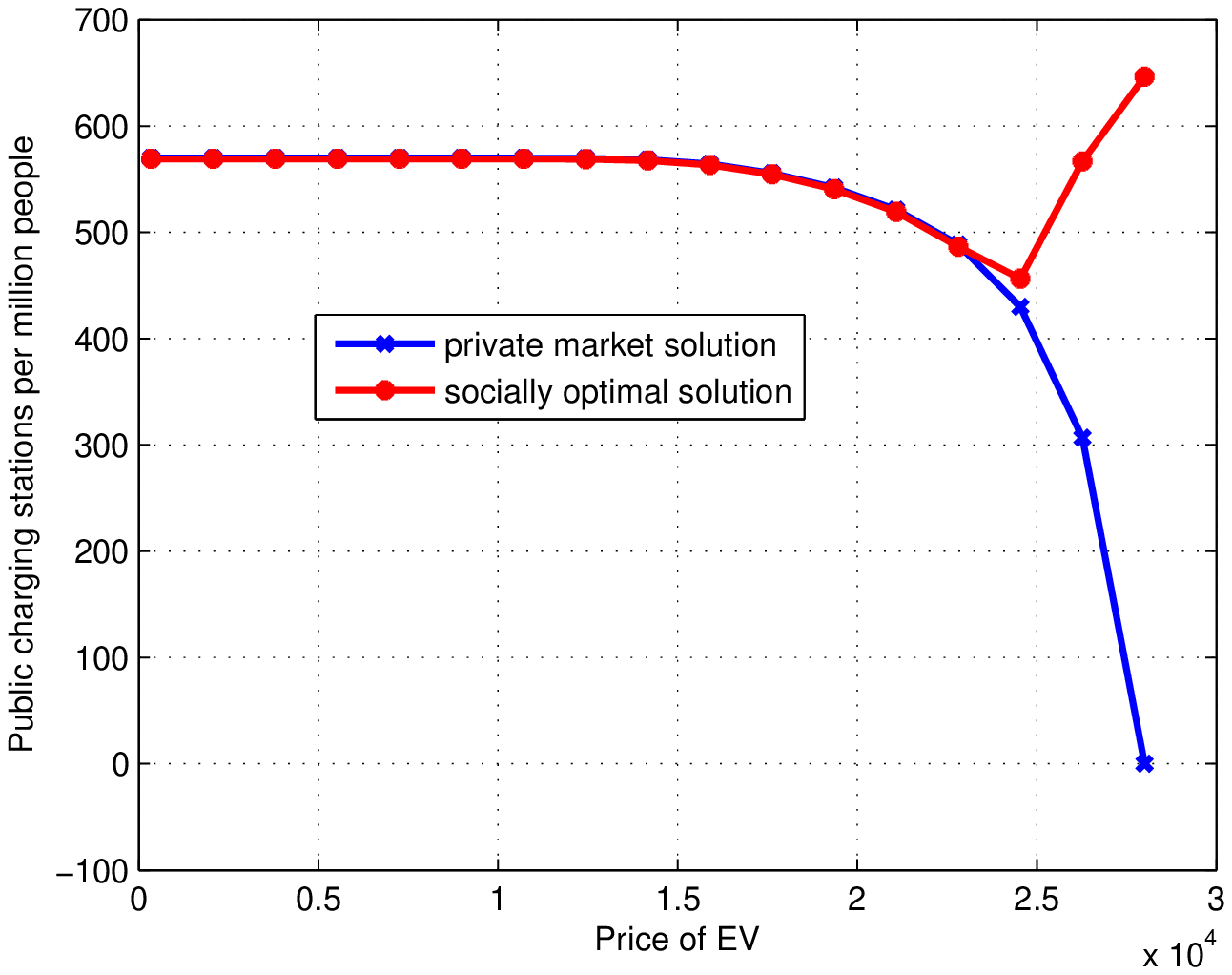}
  \end{minipage}
  }
%
\subfigure[\scriptsize Charging price vs. EV price.]{
\label{Fig:chargingPriceSW}
  \begin{minipage}{.3\textwidth}
    \includegraphics[width=1.21\textwidth]{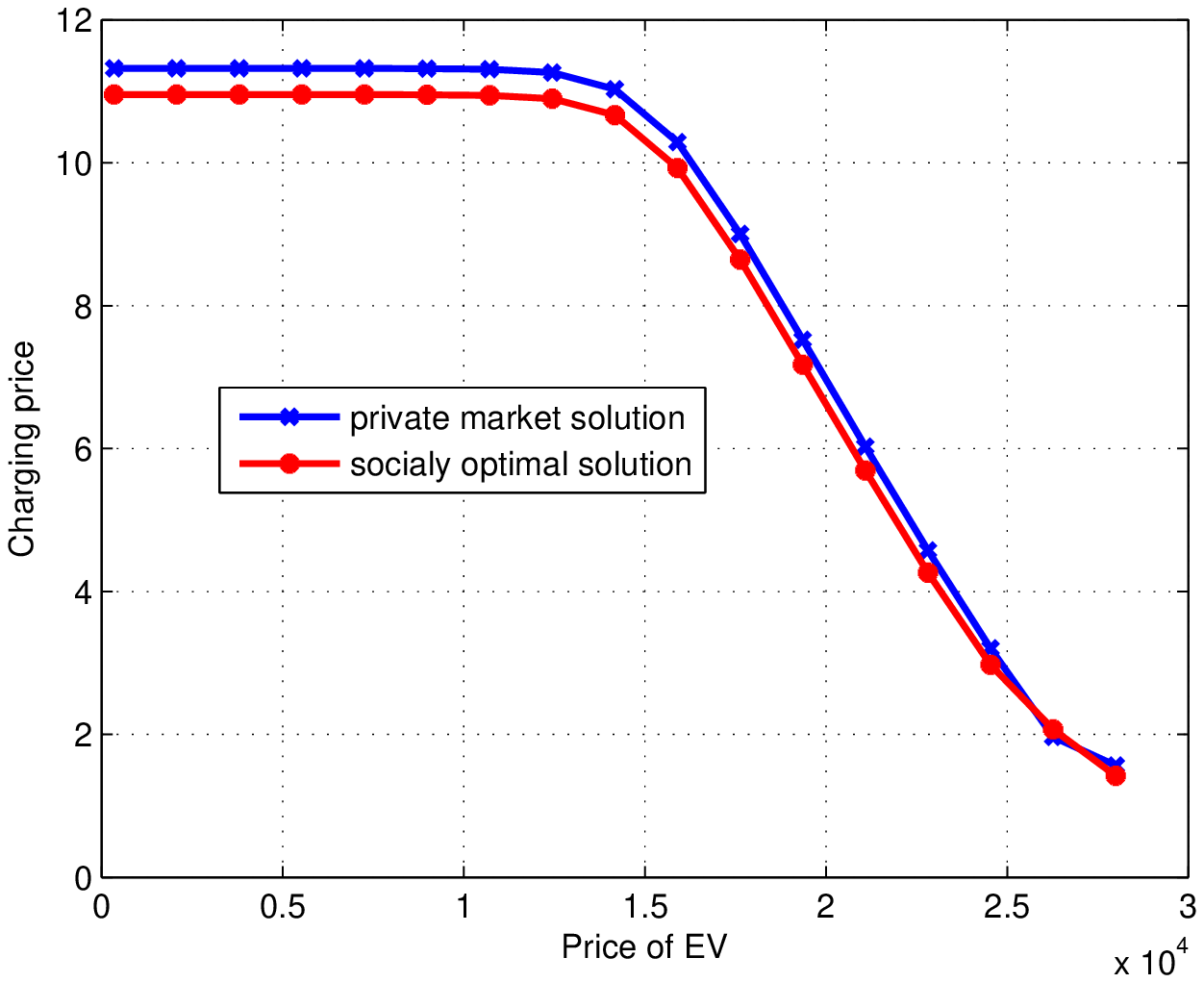}
  \end{minipage}
   }
  \subfigure[\scriptsize Probability of charging outside vs. EV price.]{
  \label{Fig:proChargingOutside}
  \begin{minipage}{.3\textwidth}
    \includegraphics[width=1.21\textwidth]{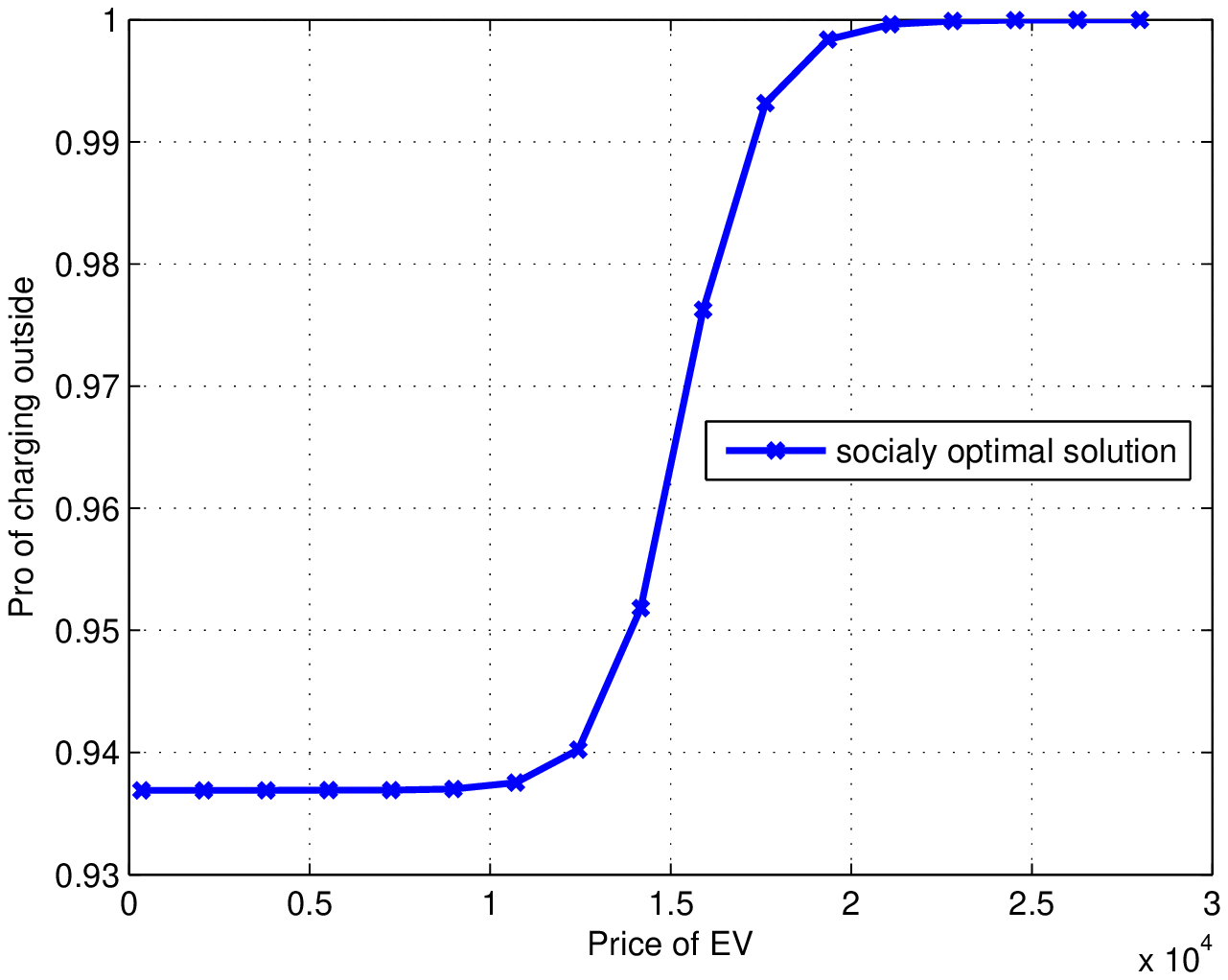}
  \end{minipage}
  }
  \subfigure[\scriptsize Charging utility vs. EV price.]{
  \label{Fig:chargingUtility}
  \begin{minipage}{.3\textwidth}
    \includegraphics[width=1.21\textwidth]{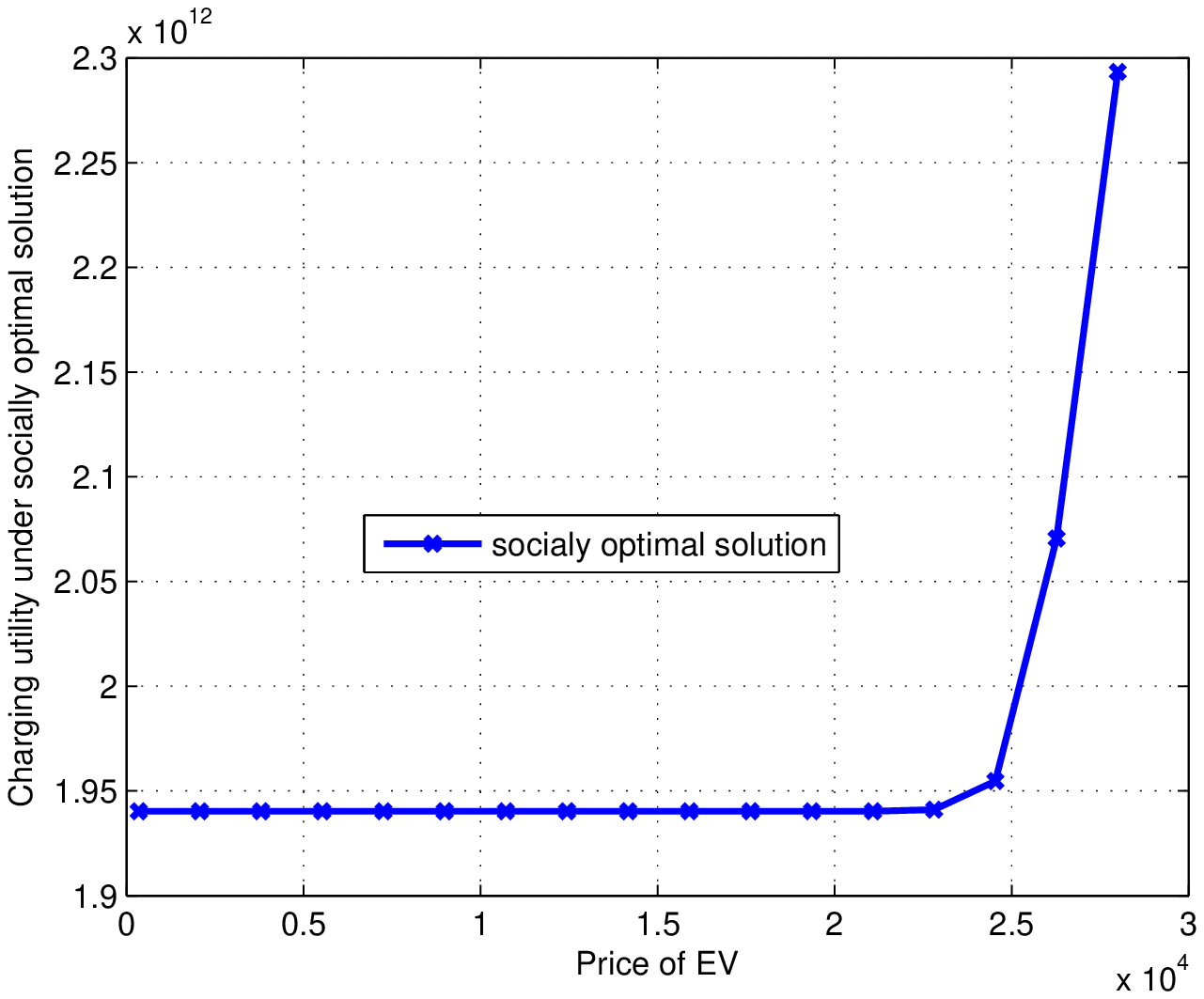}
  \end{minipage}
  }
  \centering
  \caption{Socially optimal solution vs. private market solution}
\end{figure}

As shown in Fig~\ref{Fig:marketShareSW}, when the EV price is low, both solutions result in higher EV market shares. When the EV price increases, the EV market share from a socially optimal solution drops slower than from a private market solution. We can gain insights into this from Fig.~\ref{Fig:densitySW}. When the EV price is low, both solutions lead to a large density of EVCSs. When the EV price increases, the charging demand shrinks because the EV market share drops. In the private market solution, less charging facilities will be built due to the decreasing operational profit. However, the socially optimal solution mandates a presence of higher level of EVCSs. More EVCSs attract more consumers to purchase EVs thus slow down the drop of the EV market share. As shown in Fig~\ref{Fig:chargingPriceSW}, the charging price also drops with the price of EV. Lower charging price attracts more consumers to charge at the public charging stations as shown in Fig~\ref{Fig:proChargingOutside}. As a result of low charging price, the total charging utility of consumers increases dramatically and dominates the profit of charging stations. The social planner need to build more charging stations to maintain the utility of consumers, which explained the trend change in Fig.~\ref{Fig:densitySW}. This is an intuitive result that at the EV launch stage, the social planner should balance the interests of the EVCS investor and that of the consumer.


\section{Conclusion}\label{sec:VI}
In this paper, the two-sided market problem of EV-EVCS is considered. A sequential game is formulated to analyze the indirect network effects between the charging station investor and consumers. The optimal operation decision of charging stations is shown as locational equal profit pricing. An asymptotic optimal algorithm of investment decision is proposed which reduces the computation complexity significantly. The social welfare optimization is discussed and it is shown that the socially optimal solution requires more charging stations than the market outcomes. The numerical results are presented to illustrate the impact of government subsidy policy and the difference between the private market solution and the socially optimal solution.

As an analytical approach to understanding the market dynamics of EV diffusion, this paper assumes a stylized model for both the consumers and the investor.  Here we aim to capture major factors in the interactions between the consumers and the investor, including the EV price, the coverage of the charging stations, and the price of charging. Ignored in the model includes several nontrivial and practically significant factors.  For instance, the price of EV is assumed exogenous, and the EV consumers and charging stations are mostly homogeneous (except that the favorability rating, the operating cost, and pricing of charging are different across locations).  A multi-stage counter part of this work is to be reported in the future.  

\section*{Appendix}\label{sec:VII}
\numberwithin{equation}{section}
\appendix
Here the proofs under the assumption of the type I extreme value vehicle preference distribution are presented. The uniform preference distribution proof can be found in~\citep{Yu&Li&Tong:14Allerton}. The subscript $j$ is dropped in this section for convenience.

\section{Proof of Theorem~\ref{thm:priceConvergence}}\label{sec:priceConvergence}
In Theorem~\ref{thm:optimalPrice}, the optimal charging price is shown to generate uniform profits across charging stations. Denote the uniform profit by ${r\triangleq\rho^*_i-c_i}$, the sum of the exponential systematic utility by ${v=\sum_{i=1}^{N_E}v_i=\sum_{i=1}^{N_E}\exp(\alpha_1f_i-\alpha_2c_i)}$ and the ratio of the utility and the exponential profit by ${\kappa=v/\exp(\alpha_2r)}$. We first show the following result which is used to prove the theorem:

\begin{lem}\label{lem:marketShareConvergence}
\beq
\begin{array}{ll}
\lim_{v\rightarrow+\infty}\eta=\lim_{v\rightarrow+\infty}\frac{(q_0+\kappa)^{\beta_1}}{(q_0+\kappa)^{\beta_1}+C}&=1\\
\lim_{v\rightarrow+\infty}P_0=\lim_{v\rightarrow+\infty}\frac{q_0}{q_0+\kappa}&=0.\\
\lim_{v\rightarrow+\infty}r=+\infty
\end{array}
\eeq
\end{lem}

\begin{proof}
 Equation (\ref{eqn:optimalPrice_Extreme}) can be rewritten as
\beq\label{eq:firstOrderDerivative}
\begin{array}{ll}
g(v,r)&\triangleq\alpha_2\beta_1r(1-\eta)(1-P_0)+\alpha_2rP_0-1\\
&=\beta_1\ln(v/\kappa)\frac{C}{(q_0+\kappa)^{\beta_1}+C}\frac{\kappa}{q_0+\kappa}+\ln(v/\kappa)\frac{q_0}{q_0+\kappa}-1\\
&=0,
\end{array}
\eeq
where ${C=e^{[\beta_1\mathbb{E}(U_G)-\beta_2p_G+\beta_Gp_E]}}$ and ${q_0=e^{(\alpha_1f_0-\alpha_2\rho_0)}}$.

Note the first term in the first row of (\ref{eq:firstOrderDerivative}), ${\alpha_2\beta_1r(1-\eta)(1-P_0)}$, is positive. We can conclude that as $v\rightarrow+\infty$, $\kappa\rightarrow+\infty$. Otherwise, the second term $\ln(v/\kappa)\frac{q_0}{q_0+\kappa}\rightarrow+\infty$ which violates (\ref{eq:firstOrderDerivative}).

Since $q_0$ and $C$ are constant, $\beta_1>0$ and ${\kappa\rightarrow+\infty}$, ${\eta\rightarrow1}$ and ${P_0\rightarrow0}$. So the uniform profit ${r\rightarrow+\infty}$ as $v\rightarrow+\infty$. Otherwise (\ref{eq:firstOrderDerivative}) does not hold any more.
\end{proof}

Lemma \ref{lem:marketShareConvergence} is intuitive. As the sum of exponential systematic utility $v$ increases, the charging service at the EVCSs is more convenient, thus more consumers tend to purchase EVs and prefer to charge at the public charging stations.

Now we show the proof of Theorem \ref{thm:priceConvergence} using Lemma \ref{lem:marketShareConvergence}.
\begin{proof}
By applying the implicit function theorem (IFT) to function $g(v,r)$, we get the derivative of profit $r$ with respect to the exponential systematic utility $v$,
\beq\label{eq:profitDerivative}
\frac{\partial r}{\partial v}=-\frac{\partial g(v,r)/\partial v}{\partial g(v,r)/\partial r}=\frac{1}{\alpha_2v}\frac{1}{1+h},
\eeq
where
\scriptsize
\[
h=\frac{\beta_1(1-\eta)(1-P_0)+P_0}{\alpha_2\beta_1^2
r\eta(1-\eta)(1-P_0)^2-\alpha_2\beta_1r(1-\eta)P_0(1-P_0)+\alpha_2rP_0(1-P_0)}.
\]
\normalsize

By Lemma \ref{lem:marketShareConvergence}, the numerator of $h$, $\beta_1(1-\eta)(1-P_0)+P_0$, converges to $0$ as $v\rightarrow+\infty$ . Denote the denominator of $h$ by $H$. If $\beta_1>1$, $H$ can be rewritten as
\small
\beq\label{eq:firstCondition}
\begin{array}{rcl}
H&=&\alpha_2\beta_1r(1-\eta)(1-P_0)(1-P_0)+\alpha_2rP_0(1-P_0)\\
&&+\alpha_2\beta_1r(1-\eta)(1-P_0)[\beta_1\eta(1-P_0)-1]\\
&=&(1-P_0)+\alpha_2\beta_1r(1-\eta)(1-P_0)[\beta_1\eta(1-P_0)-1],
\end{array}
\eeq
\normalsize
where the last equality is because of (\ref{eq:firstOrderDerivative}). In (\ref{eq:firstCondition}), the first term, ${1-P_0}$, converges to $1$ and the second term is positive as $v\rightarrow+\infty$.

If $\beta_1\le1$, $H$ can be rewritten as
\small
\beq\label{eq:secondCondition}
\begin{array}{rcl}
H&=&\alpha_2\beta_1r(1-\eta)(1-P_0)[\beta_1\eta(1-P_0)-P_0]\\
&&+\alpha_2rP_0[\beta_1\eta(1-P_0)-P_0]+\alpha_2rP_0[1-\beta_1\eta(1-P_0)]\\
&=&[\beta_1\eta(1-P_0)-P_0]+\alpha_2rP_0[1-\beta_1\eta(1-P_0)],
\end{array}
\eeq
\normalsize
where the last equality is because of (\ref{eq:firstOrderDerivative}). In (\ref{eq:secondCondition}), the first term converges to $\beta_1$  as $v\rightarrow+\infty$ and the second term is positive.

In both cases, $H$ is bounded below from zero as $v\rightarrow+\infty$. Thus $h\rightarrow0$ and $v\frac{\partial r}{\partial v}\rightarrow\frac{1}{\alpha_2}$ in (\ref{eq:profitDerivative}).

Since $r\rightarrow+\infty$ and $\partial r/\partial v\rightarrow 0$ as $v\rightarrow+\infty$, by the l'H\^opital's rule we have
\beq
\lim_{v\rightarrow+\infty}\frac{r}{\ln v}=\lim_{v\rightarrow+\infty}\frac{\partial r/\partial v}{1/v}=\lim_{v\rightarrow+\infty}\frac{\partial^2 r/\partial v^2}{- v^{-2}}=\frac{1}{\alpha_2},
\eeq
which completes the proof.

\end{proof}

Note when $v>\bar{v}_1$ for some $\bar{v}_1>0$, $h>0$. $r(v)$ is strictly increasing in $v$ and $v\frac{\partial r}{\partial v}<\frac{1}{\alpha_2}$ when $v>\bar{v}_1$ .

\section{Proof of Theorem~\ref{thm:asymptoticOptimal}}\label{sec:ProofAsy}
The proof is divided into two parts. First, fixing the number of charging stations to build, we examine where to build these stations. Then we look into how many charging stations should be built.

We first show the following result which is used to prove the theorem:
\begin{lem}\label{lem:where}
Fixing the number of stations to build as $N_E$, the asymptotically optimal strategy of building is to pick $N_E$ candidates with largest $v_i=\exp(\alpha_1 f_i-\alpha_2 c_i)$.
\end{lem}
\begin{proof}
Denote the sum of exponential systematic utility by $v=\sum_{i=1}^{N_E}v_i$ and the uniform charging profit by $r=\rho^*_i-c_i$. The operational profit of the investor can be stated as
\beq
\Pi(v)=r(v)\eta(v)\sum_{i=1}^{N_E}P_i(v)=r(v)\frac{[q_0+\kappa(v)]^{\beta_1}}{[q_0+\kappa(v)]^{\beta_1}+C}\frac{\kappa(v)}{q_0+\kappa(v)},
\eeq
where ${q_0=\exp(\alpha_1f_1-\alpha_2\rho_0)}$, ${\kappa(v)=v\exp(-\alpha_2r)}$, and ${C=\exp(\beta_1\mathbb{E}(U_G)-\beta_2p_G+\beta_Gp_E)}$. We will show that $\Pi(v)$ is asymptotically an increasing and concave function of $v$.

The derivative of $\kappa(v)$ with respect to $v$ is stated as
\beq
\frac{\partial \kappa(v)}{\partial v}=\exp(-\alpha_2r)(1-\alpha_2v\frac{\partial r}{\partial v})
\eeq
which is strictly positive when $v>\bar{v}_1$ according to the discussion in ~\ref{sec:priceConvergence}. So the operational profit $\Pi(v)$ is strictly increasing in $v$ when $v>\bar{v}_1$.

The second order derivative of $\Pi(v)$ with respect to $v$ is stated as
\small
\beq\label{eq:secondOrderDerivative}
\begin{array}{ll}
\frac{\partial^2\Pi}{\partial v^2}=&\frac{\eta}{v^2}\{v^2\frac{\partial^2r}{\partial v^2}(1-P_0)\\
&+(-2rP_0)(1-P_0)^2\\
&+\frac{2}{\alpha_2}\frac{1}{1+h}(1-P_0)[(1-\eta)\beta_1(1-P_0)+P_0]\\
&+r(1-\eta)\beta_1(1-P_0)^2[\beta_1(1-2\eta-P_0)+(2\eta+3)P_0-1]\}.
\end{array}
\eeq
\normalsize
As $v\rightarrow+\infty$, $v^2\frac{\partial^2r}{\partial v^2}\rightarrow-\frac{1}{\alpha_2}$, $\eta\rightarrow+\infty$ and $P_0\rightarrow0$. So when $v>\bar{v}_2$ for some $\bar{v}_2>0$, the first term in (\ref{eq:secondOrderDerivative}) is negative and bounded above from zero. The second and last term are negative. The third term is positive but converging to zero. The second order derivative of $\Pi(v)$ is negative and we have the concavity of $\Pi(v)$.

The monotonicity of $\Pi(v)$ implies that, if given two station candidates $j$ and $j'$, fixing the other $(N_E-1)$ stations, the one with larger $v_i=\exp(\alpha_1 f_i-\alpha_2c_i), i\in\{j,j'\}$ should be built because the building costs are the same. So we have the optimal strategy about where to build stations as follows.
\end{proof}

\begin{figure}[h!]

  \begin{minipage}{.49\textwidth}
  \subfigure[\small Profit and cost of charging stations.]{
\label{fig:profit}
\psfrag{q}{{\small$v$}}
\psfrag{qq}{{\small$v^*$}}
\psfrag{q1}{{\small$v_0+v_1$}}
\psfrag{q2}{{\small$v_0+v_1+v_2$}}
\psfrag{q3}{{\small$v_0+v_1+v_2+v_3$}}
\psfrag{f}{{\small$\tilde{F}(v)$}}
\psfrag{NC/A}{{\small $\frac{2\phi }{\beta}$}}
\psfrag{Pi}{{\small$\Pi(v)$}}
\psfrag{high}{{\small$C_{\mbox {high}}$}}
\psfrag{proLowHigh}{{\small$1-P_{\mbox {low}}$}}
\psfrag{proHighLow}{{\small$1-P_{\mbox {high}}$}}
\psfrag{time}{{\small \mbox{time}}}
  \includegraphics[width=1\textwidth]{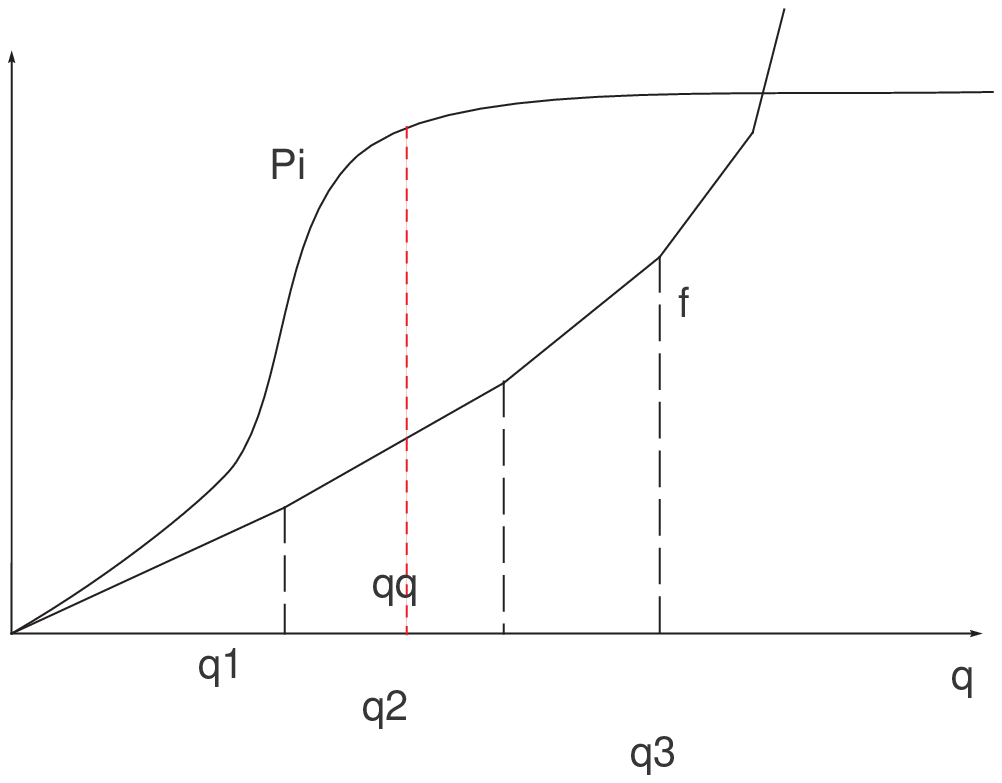}
  
   }\end{minipage}
\subfigure[\small Derivatives of profit and cost of charging stations.]{
\label{fig:derivative}
\begin{minipage}{.49\textwidth}
\psfrag{q}{{\small$v$}}
\psfrag{qq}{{\small$v^*$}}
\psfrag{q1}{{\small$v_0+v_1$}}
\psfrag{q2}{{\small$v_0+v_1+v_2$}}
\psfrag{q3}{{\small$v_0+v_1+v_2+v_3$}}
\psfrag{f}{{\small$\frac{\partial \tilde{F}(v)}{\partial v}$}}
\psfrag{NC/A}{{\small $\frac{N_C}{\alpha_2}$}}
\psfrag{Pi}{{\small$\frac{\partial\Pi(v)}{\partial v}$}}
\psfrag{high}{{\small$C_{\mbox {high}}$}}
\psfrag{proLowHigh}{{\small$1-P_{\mbox {low}}$}}
\psfrag{proHighLow}{{\small$1-P_{\mbox {high}}$}}
\psfrag{time}{{\small \mbox{time}}}
  \includegraphics[width=1.2\textwidth]{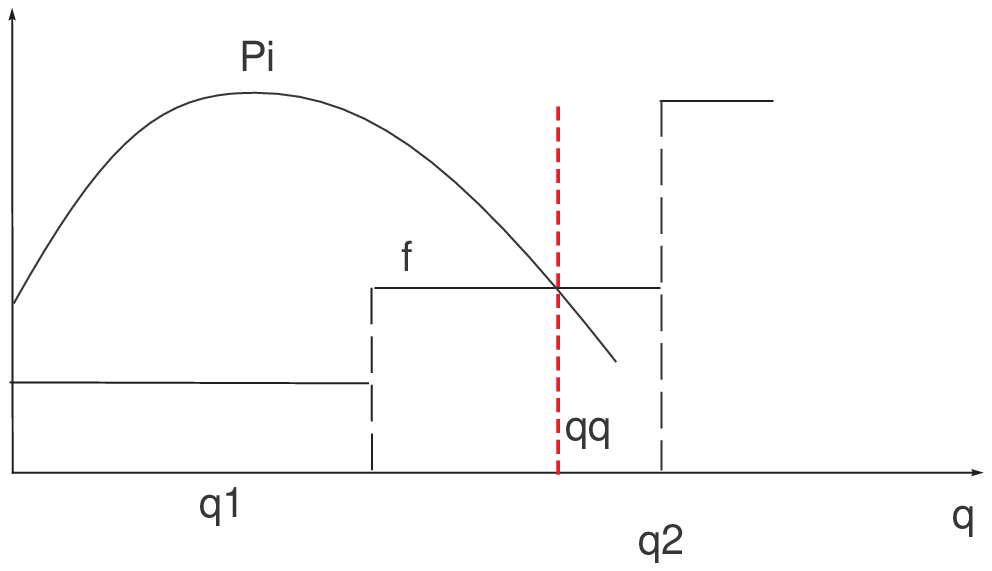}
  \end{minipage}
  }
  \centering
  \caption{Trends of profit and cost functions of charging stations.}
\end{figure}

Knowing where to build charging stations, we show the proof of the optimality of the greedy algorithm.
\begin{proof}
After we sort the $N_L$ candidate locations by $v_i$, we can present the cost ${\sum_{i=1}^{N_E}F(s_i)=(1+\gamma)F_0N_E}$ as a function of $v=\sum_{i=1}^{N_E}v_i$. We can treat $v$ as a continuous variable and write ${\tilde{F}(v)\triangleq(1+\gamma)F_0N_E}$ as ${\tilde{F}(v)=(1+\gamma)F_0N+(v-\sum_{i=1}^{N}v_i)(1+\gamma)F_0/v_{N+1}}$, if ${\sum_{i=1}^{N}v_i< v\le \sum_{i=1}^{N+1}v_i}$.
Since $v_i\ge v_{i+1}$, the cost $\tilde{F}(v)$ is a piece wise linear convex function of $v$. The partial derivative $\partial\tilde{F}(v)/\partial v$ is piece wise constant and increasing in $v$.

The trends of $\Pi(v)$, $\tilde{F}(v)$ and the derivatives are plotted in Fig.~\ref{fig:profit} and~\ref{fig:derivative}. In Fig.~\ref{fig:profit}, $v^*$ is the optimal point to maximize the profit $(\Pi(v)-\tilde{F}(v))$. Fig.~\ref{fig:derivative} shows the derivative of $\tilde{F}(v)$ is increasing and the marginal profit $\frac{\partial\Pi(v)}{\partial v}$ is decreasing when $v$ is large enough. The last cross point of the derivatives of $\Pi(v)$ and $\tilde{F}(v)$ is the optimal point.
Combining Lemma ~\ref{lem:where}, we have the asymptotic optimality.

\end{proof}

%

\section{Proof of Theorem~\ref{thm:socialWelfare}}\label{proof:socialWelfare}
\begin{proof}
    Denote the sum of consumer utility and investor's operational profit by ${\bar{S}_W(v)=S_C(v)+\Pi(v)}$, the social planner is maximizing $(\bar{S}_W(v)-\tilde{F}(v))$.

    The consumer utility, $S_C(v)$, can be rewritten as a function of the total exponential systematic utility $v$ as follows.
    \[
    S_C(v)=\ln\Big[\Big(q_0+v\exp(-\alpha_2r)\Big)^{\beta_1}C_1+C_2\Big],
    \]
    which is increasing in $v$. So
    \[
    \frac{\partial \bar{S}_W(v)}{\partial v}=\frac{\partial S_C(v)}{\partial v}+\frac{\partial \Pi(v)}{\partial v}>\frac{\partial \Pi(v)}{\partial v}.
    \]
    We plot the derivative of the social welfare as well as that of the investor utility in Fig~\ref{fig:socialWelfare}.
     \begin{figure}
\centering
\psfrag{q}{{\small$v$}}
\psfrag{qq}{{\small$v^*$}}
\psfrag{qqq}{{\small$v^{**}$}}
\psfrag{q1}{{\small$v_0+v_1$}}
\psfrag{q2}{{\small$v_0+v_1+v_2$}}
\psfrag{q3}{{\small$v_0+v_1+v_2+v_3$}}
\psfrag{f}{{\small$\frac{\partial \tilde{F}(v)}{\partial v}$}}
\psfrag{si}{{\small $\frac{\partial \Pi(v)}{\partial v}$}}
\psfrag{Pi}{{\small$\frac{\partial \bar{S}_w(v)}{\partial v}$}}
\psfrag{high}{{\small$C_{\mbox {high}}$}}
\psfrag{proLowHigh}{{\small$1-P_{\mbox {low}}$}}
\psfrag{proHighLow}{{\small$1-P_{\mbox {high}}$}}
\psfrag{time}{{\small \mbox{time}}}
  \includegraphics[width=0.6\textwidth]{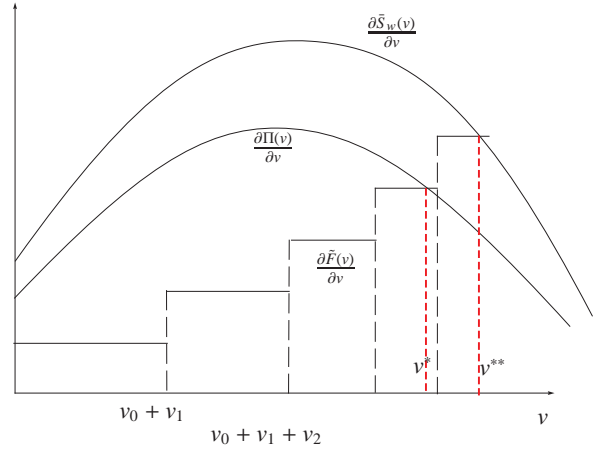}\\
  \caption{Derivatives of social welfare and investor utility.}
  \label{fig:socialWelfare}
\end{figure}
The optimal social welfare point $v^{**}$ is also the cross point of $\frac{\partial \tilde{F}(v)}{\partial v}$ and $\frac{\partial \bar{S}_W(v)}{\partial v}$. Since $\frac{\partial \bar{S}_W(v)}{\partial v}>\frac{\partial \Pi(v)}{\partial v}$, it is always true that $v^{**}\ge v^{*}$, which implies the socially optimal solution requires more charging stations than the private market outcomes.
\end{proof} 

\section*{Acknowledgement}
This work is supported in part by the National Science Foundation under Grant CNS-1248079.
\bibliographystyle{elsarticle-harv}
\bibliography{Bibs/Journal,Bibs/Conf,Bibs/Book,Bibs/Misc}







\end{document}